\documentclass[twocolumn]{aastex63}
\usepackage{graphics,graphicx}
\hypersetup{urlcolor=blue}
\usepackage{natbib}
\usepackage[outdir=./]{epstopdf}
\usepackage{textcomp,gensymb}
\usepackage{xfrac}
\usepackage{xcolor}
\usepackage{multirow}

\newcommand{\kepler}{{\it Kepler}}

\newcommand{\um}{$\mu$m}
\newcommand{\fbol}{$F_{\mathrm{bol}}$}

\newcommand{\teff}{\ensuremath{T_{\text{eff}}}}

\newcommand{\gaia}{{\textit Gaia}}

\usepackage[outdir=./]{epstopdf}
\usepackage{ulem}
\usepackage{mathrsfs}
\usepackage{fontawesome}
\usepackage{comment}
\usepackage[version=4]{mhchem} 

\newcommand{\tess}{\textit{TESS}}
\newcommand{\ktwo}{{\textit K2}}
\newcommand{\mearth}{MEarth}
\newcommand{\hst}{\textit{HST}}
\newcommand{\tsurf}{$T_{\rm{surf}}$}
\newcommand{\tspot}{$T_{\rm{spot}}$}
\newcommand{\spitzer}{\textit{Spitzer}}
\newcommand{\hubble}{\textit{Hubble}}

\pdfoutput=1

\shorttitle{Hazes in K2-33b} 
\shortauthors{Thao et al. 2022}

\begin{document}

\title{Hazy with a chance of star spots:\\ constraining the atmosphere of the young planet, K2-33b} 

\correspondingauthor{Pa Chia Thao}
\email{pachia@live.unc.edu} 

\author[0000-0001-5729-6576]{Pa Chia Thao}
\altaffiliation{NSF Graduate Research Fellow}
\altaffiliation{Jack Kent Cooke Foundation Graduate Scholar}
\affiliation{Department of Physics and Astronomy, The University of North Carolina at Chapel Hill, Chapel Hill, NC 27599, USA}

\author[0000-0003-3654-1602]{Andrew W. Mann}
\affiliation{Department of Physics and Astronomy, The University of North Carolina at Chapel Hill, Chapel Hill, NC 27599, USA}

\author[0000-0002-8518-9601]{Peter Gao}
\affiliation{Earth and Planets Laboratory, Carnegie Institution for Science, 5241 Broad Branch Road, NW, Washington, DC 20015, USA}

\author[0000-0002-6397-6719]{Dylan A. Owens}
\affiliation{Department of Physics and Astronomy, The University of North Carolina at Chapel Hill, Chapel Hill, NC 27599, USA}
\affiliation{Gemini Observatory, NSF's NOIRLab, 670 N. A'ohoku Place Hilo, Hawaii, 96720, USA}

\author[0000-0001-7246-5438]{Andrew Vanderburg}
\affiliation{Department of Physics and Kavli Institute for Astrophysics and Space Research, Massachusetts Institute of Technology, Cambridge, MA 02139, USA}

\author[0000-0003-4150-841X]{Elisabeth R. Newton}%
\affiliation{Department of Physics and Astronomy, Dartmouth College, Hanover, NH 03755, USA}

\author{Yao Tang}
\affiliation{University of California, Santa Cruz, Santa Cruz, CA 95064, USA}

\author[0000-0002-9641-3138]{Matthew J. Fields}
\affiliation{Department of Physics and Astronomy, The University of North Carolina at Chapel Hill, Chapel Hill, NC 27599, USA}

\author[0000-0001-6534-6246]{Trevor J. David}
\affiliation{Center for Computational Astrophysics, Flatiron Institute, New York, NY 10010, USA} 
\affiliation{American Museum of Natural History, Central Park West, New York, NY 10024, USA}

\author{Jonathan M. Irwin}
\affil{Institute of Astronomy, University of Cambridge, Madingley Road,
Cambridge, CB3 0HA, UK}

\author[0000-0003-2466-5077]{Tim-Oliver Husser}
\affiliation{Institut für Astrophysik, Georg-August-Universität Göttingen, Friedrich-Hund-Platz 1, 37077 Göttingen, Germany}

\author[0000-0002-9003-484X]{David Charbonneau}
\affiliation{Harvard-Smithsonian Center for Astrophysics, Harvard University, Cambridge, MA 02138, USA}

\author[0000-0002-3247-5081]{Sarah Ballard}
\affiliation{Bryant Space Science Center, Department of Astronomy, University of Florida, Gainesville, FL 32611, USA}

\begin{abstract} 
Although all-sky surveys have led to the discovery of dozens of young planets, little is known about their atmospheres. Here, we present multi-wavelength transit data for the super Neptune-sized exoplanet, K2-33b -- the youngest ($\sim$10\ Myr) transiting exoplanet to-date. We combined photometric observations of K2-33 covering a total of 33 transits spanning $>$2 years, taken from \ktwo, MEarth, \hst, and \textit{Spitzer}. The transit photometry spanned from the optical to the near-infrared (0.6-4.5µm), enabling us to construct a transmission spectrum of the planet. We find that the optical transit depths are nearly a factor of two deeper than those from the near-infrared. This difference holds across multiple datasets taken over years, ruling out issues of data analysis and unconstrained systematics. Surface inhomogeneities on the young star can reproduce some of the difference, but required spot coverage fractions ($>$60\%) are ruled out by the observed stellar spectrum ($<$20\%). We find a better fit to the transmission spectrum using photochemical hazes, which were predicted to be strong in young, moderate-temperature, and large-radius planets like K2-33b. A tholin haze with \ce{CO} as the dominant gaseous carbon carrier in the atmosphere can reasonably reproduce the data with small or no stellar  surface inhomogeneities, consistent with the stellar spectrum. The \hst\ data quality is insufficient for the detection of any molecular features. More observations would be required to fully characterize the hazes and spot properties and confirm the presence of CO suggested by current data. 
\end{abstract}

\keywords{ Exoplanet atmospheres; Transit photometry; Open star clusters; Exoplanet evolution; Starspots; M dwarf stars; Markov chain Monte Carlo; Light curves; Starspots}

\section{Introduction}\label{sec:intro}

The study of exoplanet atmospheres has progressed at a tremendous pace in the last decade due to the high-sensitivity spectroscopic observations covering secondary eclipse spectroscopy, high-resolution Doppler spectroscopy, and direct imaging spectroscopy. Transmission spectroscopy, a measurement of the planet radius as a function of wavelength, has enabled the detection of a range of atomic and molecular species \citep[e.g.,][]{seager2000theoretical, brown2001transmission, hubbard2001theory, Huitson2013, Sing2015, Sing2016}. Detecting such molecules can provide information about the planet's formation and migration history, although complexities in the planet and disk properties can make such detections difficult to interpret. For example, planets formed beyond the snow line will have an atmosphere that is carbon rich (an enhanced C/O) compared to planets formed closer-in \citep{Oberg2011, booth2017chemical, booth2019planet}.

Measurements of young exoplanet atmospheres offer a more direct correlation to the disk, which is unadulterated by later-stage evolution. However, the majority of planets that have had their atmosphere characterized are mature ($>$1 Gyr), or have unconstrained ages. This is because young planetary systems with precise ages are rare \citep[usually limited to those in young associations, e.g.,][]{Mann2016a, Newton2019, Bouma2020, David2019, Obermeier2016} and are harder to detect due to stellar variability. Only in the last decade has there been a significant increase in the number of known young planetary systems. This has been driven by improvements in near-infrared radial velocities \citep[e.g.,][]{2016ApJ...826..206J}, mitigation of stellar variability \citep[e.g.,][]{Donati:2017aa, Rizzuto2017}, as well as, NASA's \ktwo\ \citep{Howard2014} and \tess\ \citep[Transiting Exoplanet Survey Satellite;][]{Ricker2014} missions surveying nearby young clusters and associations.

Only three young ($<$1 Gyr) exoplanets have had their atmospheres characterized through transmission spectroscopy: \footnote{While there are several young planets \citep[$\beta$ Pictoris b, 51 Eridani b, HR 8799 b;][]{Chilcote2017, Samland2017, Lee2013} with direct spectroscopy, these are more massive and at wider separations ($\gg$ 1 AU) than the planets transits survey; they likely have a different formation history.}  K2-25b, Kepler 51b, and Kepler 51d -- all of which are adolescent ($\approx$ 500-700 Myr) and may be in the midst of undergoing thermal contraction. Statistical analyses of \textit{Kepler} planet candidates suggest that most planets larger than 1.6 R$_{\oplus}$ -- as is the case for the aforementioned planets -- have an extended envelope with a low mean molecular weight (e.g., of H/He) \citep{Rogers2015}, and therefore were predicted to have a large scale height, resulting in strong spectral features. However, all three planets showed a featureless transmission spectra \citep{Thao2020, Libby-Roberts2020}, which were interpreted as ongoing atmospheric mass loss (expected in such young systems) leading to dusty atmospheric outflows with small dust grains \citep{Wang2019} or high-altitude photochemical hazes \citep{Gao2020}. The small particle size leads to a lower opacity (and hence shallower transit) in the NIR compared to the optical. Grains or hazes also inflate observed radii \citep{lammer2016identifying, cubillos2017overabundance} and weaken the spectral features. \citet{Gao2020} showed that this effect is strongest on young, low-mass planets with a moderate or low equilibrium temperature. Additional objects and broader wavelength coverage could confirm these predictions.

A reoccurring challenge in interpreting transmission spectra is the contamination by star spots and faculae \citep{rackham2017access}. Spots can change the observed signal, whether or not the planet crosses them. In the case of unocculted spots, the transit chord will be brighter than the stellar average, leading to a deeper transit with increasing spot coverage and contrast even for fixed planet size. Since spot contrast depends on wavelength, such unocculted spots can introduce features and  produce haze-like variations in the transmission spectrum that are not present in the planet. \citet{Barclay2021} showed that spots alone can explain the observed H$_2$O features seen in the transmission spectrum of K2-18\,b \citep{benneke2019water, tsiaras2019water}. The imprint of the star on the planet's spectrum is particularly concerning for young planets, as spot coverage fractions are statistically higher for young stars \citep{Morris2020, Luger2021}. 

\begin{figure}[htp]
    \centering
    \includegraphics[width=\columnwidth]{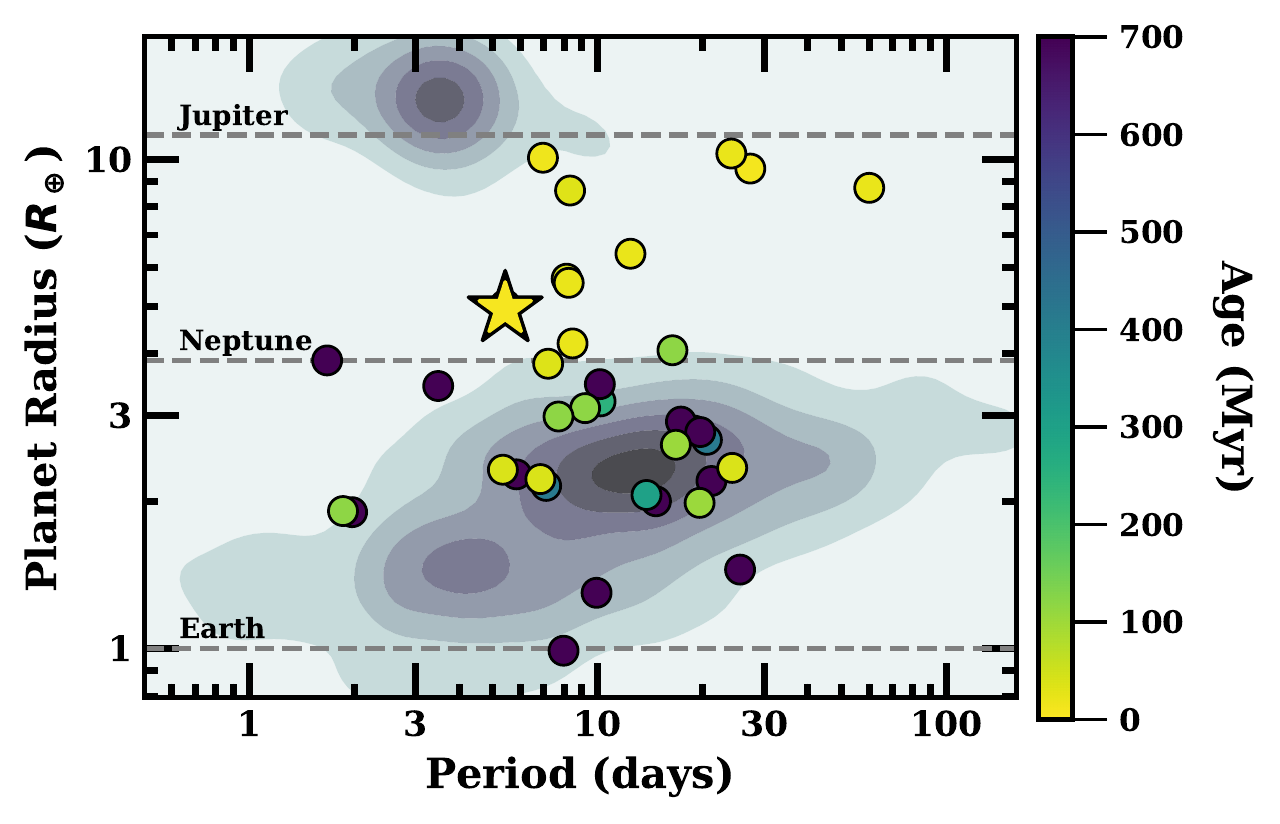}
    \caption{The detected population of all planets from \kepler\ and \ktwo\ as a function of planet radius ($R_{\oplus}$) and orbital period (days). Young ($<$1 Gyr), transiting planets that are members of clusters or stellar associations are plotted in circles colored by their age. K2-33b is outlined as a star and lands at the edge of the sub-Saturn desert. The youngest planets (yellow) have radii between Neptune and Jupiter, while no mature planets (purple) are located within this region. K2-33b is likely still evolving and offers an opportunity to observe an atmosphere in transition. Planet properties from \citet{NASAexplanetarchive}. \label{fig:young_planets}}
\end{figure}

To understand the evolution and formation of planetary atmospheres, a greater sample size of young planets -- particularly new-born planets -- need their atmospheres characterized. To this end, we explore the atmospheric transmission spectrum of the youngest transiting exoplanet to-date, K2-33b \citep{David2016, Mann2016b}. This $\simeq$ 10\,Myr, super Neptune-sized exoplanet orbits a late type pre-main sequence M3.5 dwarf in the Upper Scorpius OB Association every 5.42 days. Its youth, moderate equilibrium temperature ($\sim$770 K) and abnormally large radius (5$R_\oplus$ in the optical wavelengths; Figure~\ref{fig:young_planets}) make it an ideal target for a strong haze detection. We combined 14 transits from the discovery \ktwo\ data with 7 transits taken from the MEarth survey, 10 transits taken from \textit{Spitzer}, and a partial transit taken with \textit{HST}. 

The paper is presented as follows: Section~\ref{sec:obs} describes our observations and data reduction. Using the precise parallax from \textit{Gaia}, we update K2-33's stellar parameters in Section~\ref{sec:stellar_params}. We utilize this information in our fit to the transit light curve, as described in Section~\ref{sec:transit_fitting}. Due to the planet's youth, we discuss the effect of surface inhomogeneities on the transmission spectrum in Section~\ref{sec:transit_spots}. In Section~\ref{sec:hazes}, we investigate photochemical hazes inferred from the transmission spectrum of K2-33b. In Section~\ref{sec:summary}, we conclude with a brief summary of our results.

\section{Observation and Data Reduction}\label{sec:obs}

We collected 33 total transits of K2-33b obtained from 2014 to 2017 taken by \ktwo, the \mearth\ Project, the \spitzer\ \textit{Space Telescope}, and the \hubble\ \textit{Space Telescope} (\textit{HST}). The combined data sets span from the visible to the near-infrared (0.64-4.5$\mu$m). The details of each dataset are summarized in Table \ref{tab:obslog}.

\subsection{\ktwo}
The \ktwo\ light curve covered 14 transits from 2014 August 23 to 2014 November 13 (UT). We used the \ktwo\ data with extraction and reduction as described in one of the the discovery paper \citep{Mann2016a}. To briefly summarize, following \citet{Vanderburg2016} and \citet{Mann2016a}, we derived a correction to the raw light curve using least-square minimization to simultaneously fit for low frequency variations from stellar activity, the \kepler\ flat field \citep{Vanderburg2014}, and the transits of K2-33b. As shown in \citet{Grunblatt:2016aa}, this simultaneous fit was required to avoid biasing both the flat-field correction and the resulting transit depth; the bias is larger for highly variable stars like K2-33. 

Unlike \citet{Mann2016a}, we did not apply our initial least-squares stellar variability fit before passing it to our transit analysis (Section~\ref{sec:transit_fitting}). We instead included the final stellar variability fit as part of the MCMC (Markov chain Monte Carlo) analysis in the form of a periodic GP (Gaussian Process) kernel (Section~\ref{sec:transit_fitting}). While the initial variability fit is consistent with that derived from our transit analysis, the stellar signal is large enough on transit timescales that the earlier fit may yield underestimated errors on the transit parameters.

\begin{figure*}[htp]
    \centering
    \includegraphics[trim=0cm 53cm 0cm 0cm, clip=True,width=0.9\textwidth]{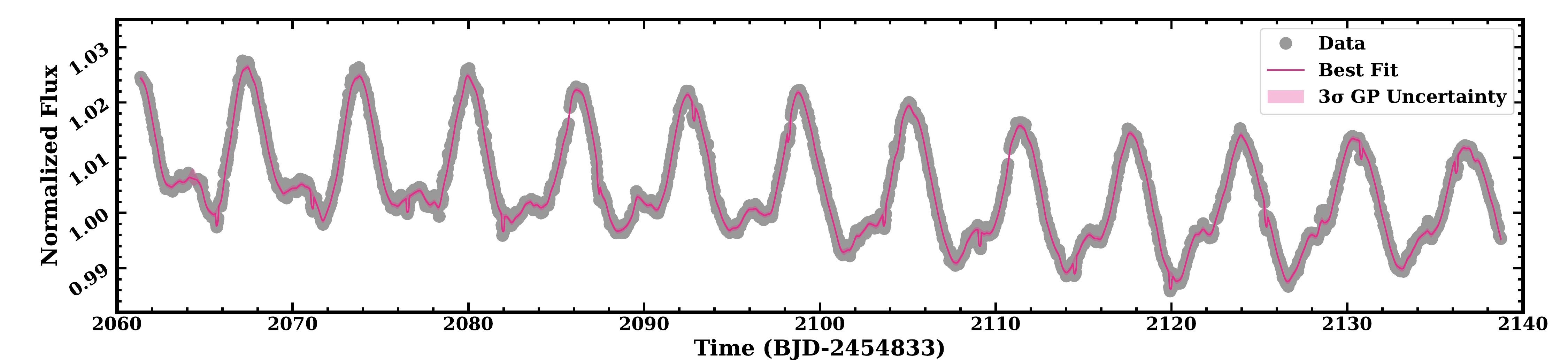}
    \caption{\ktwo\ light curve of K2-33 after removing signals from the pointing drift of the telescope (grey points). The pink shaded region shows the 3$\sigma$ range of fits from the GP variability fit, while the pink solid line shows the combined GP and the transit model. 
    \label{fig:k2_with_gp}}
\end{figure*}

\subsection{\mearth}
Including the two transits discussed in one of the discovery paper, \cite{Mann2015b}, we observed a total of nine transits (or partial transits) of K2-33b using the \mearth-North and/or \mearth-South arrays \citep{Nutzman2008, 2015csss...18..767I} spanning 2016 February 16 to 2017 June 13 (UT). Due to differences in weather and target visibility, some transits were observed with one of the \mearth\ arrays, and some transits were observed with both \mearth\ arrays (see Table~\ref{tab:obslog}). 

\mearth-North includes eight 40-cm telescopes at Fred Lawrence Whipple Observatory on Mount Hopkins, Arizona, and \mearth-South uses a nearly identical set of telescopes located at Cerro Tololo Inter-American Observatory (CTIO) in Chile. All telescopes were equipped with a $2048\times2048$ pixel CCD; \mearth-North CCDs had a pixel scale of 0.78\arcsec / pixel and \mearth-South had a pixel scale of 0.84\arcsec / pixel. For all observations, we used a Schott RG715 filter \citep{2016ApJ...818..153D}. All telescopes integrated for 60\,s for a cadence of $\simeq$90\,s per telescope. 

MEarth data were reduced following the basic methodology from \citet{2007MNRAS.375.1449I} with additional steps detailed in the documentation of the fourth \mearth\ data release\footnote{\href{https://www.cfa.harvard.edu/MEarth/DR4/processing/index.html}{https://www.cfa.harvard.edu/MEarth/DR4/processing/index.html}}. Since out-of-transit baseline varied considerably between transits, we did not perform additional systematics correction to the data (e.g., for stellar variability), and instead fit such systematics with our GP regression simultaneously with the transit (see Section~\ref{sec:transit_fitting}). 

Following reduction, we decided to remove two transits from our analysis, resulting in only 7 transits. The two transits were removed due to: 1) the data taken on 2016 July 17 (UT) only had 1.5hr of usable data near ingress; 2) the data taken on 2017 May 22 (UT) showed two sudden changes in flux during the transit, which is likely due to the instrument's meridian flip and a later halt in data from \mearth-South, or a poorly-timed flare. For the latter transit, it was difficult to remove or mitigate these problems because this transit did not contain an egress and our GP model could not simultaneously fit for such effects.

\subsection{\spitzer}
We obtained 10 full transits of K2-25b, five in 3.6$\mu$m (Channel 1) and five in 4.5$\mu$m (Channel 2), taken by the InfraRed Array Camera \citep[IRAC; ][]{Fazio2004} on the \spitzer\ \textit{Space Telescope} \citep{Werner2004}. Observations were executed over the period of 2016 Nov 08 to 2017 Nov 29 (Program ID: 13037, PI: A. Mann). Two additional transits were also taken in each band for another program (Program ID: 11026, PI: M. Werner), but were not included in this analysis. All \spitzer\ data analyzed here used the $32\times32$ pixels sub-array mode, with each image taken as a 2 s exposure. Each transit consisted of a $\simeq$24\ minute dither, a $\simeq$360\ minute stare of the full transit and out-of-transit baseline, followed by another $\simeq$7\ minute dither\footnote{\url{https://irachpp.spitzer.caltech.edu/}}. The initial dither allows an initial settling time at the new pointing position. For the long stare, we used the peak-up pointing mode, which keeps the star stable on a 0.5$\times$0.5 pixel box region of the IRAC CCD with relatively uniform sensitivity \cite[the `sweet spot';][]{Ingalls2012,Ingalls2016}. 

We first processed the flat-fielded and dark-subtracted Basic Calibrate Data (BCD) images produced by the \textit{Spitzer} pipeline using the Photometry for Orbits, Eccentricities, and Transits (\texttt{POET}; \citealt{StevensonTransit2012, 2011ApJ...727..125C}) \footnote{\href{https://github.com/kevin218/POET}{https://github.com/kevin218/POET}} pipeline to create systematics-corrected light curves. The process includes masking and flagging bad pixels, and calculating the Barycentric Julian Dates for each frame. The center position of the star was fitted using a two-dimensional, elliptical Gaussian in a 15 pixel square window centered on the target’s peak pixel. Simple aperture photometry was performed using a radius of 2.0 to 4.0 pixels, in increments of 0.25 pixels, with an inner sky annulus set to 7 pixels, and an outer sky annulus set to 15 pixels. We elected to set the radius to 2.25 pixels for both channels as it minimizes the standard deviation of the normalized residuals (SDNR). Additional reduction was done simultaneously with fitting the transit and is described in Section~\ref{sec:transit_fitting}.

\subsection{\hubble}
\hst\ observed a single transit of K2-33b using the Wide Field Camera 3 (WFC3) on 2017  May 23 (ID: 14887; PI: B. Benneke), spanning six 30-min spacecraft orbits. Observations used the G141 grating, which provided a wavelength coverage of 1.1-1.7$\mu$m, and  was taken in the 256 $\times$ 256 sub-array mode to reduce overhead times with the NSAMP=7 and SAMP-SEQ=SPARS25 readout settings. In the beginning of the observations, a direct (non-dispersed) image of K2-33 was taken in the F139M Filter for calibration purposes. Afterwards, a total of 95 spectroscopic images were taken in the forward direction of the spatial-scanning mode, in which the stellar spectrum is spread along the spatial direction to avoid non-linearity or saturation. All spectra were taken with a scan rate of 0.037 arcseconds/second for an effective exposure time of $\sim$112 seconds to produce a scan 4.38 arcseconds in length. 

\begin{figure}[ht]
    \centering
    \includegraphics[width=0.48\textwidth, height=0.25\textwidth]{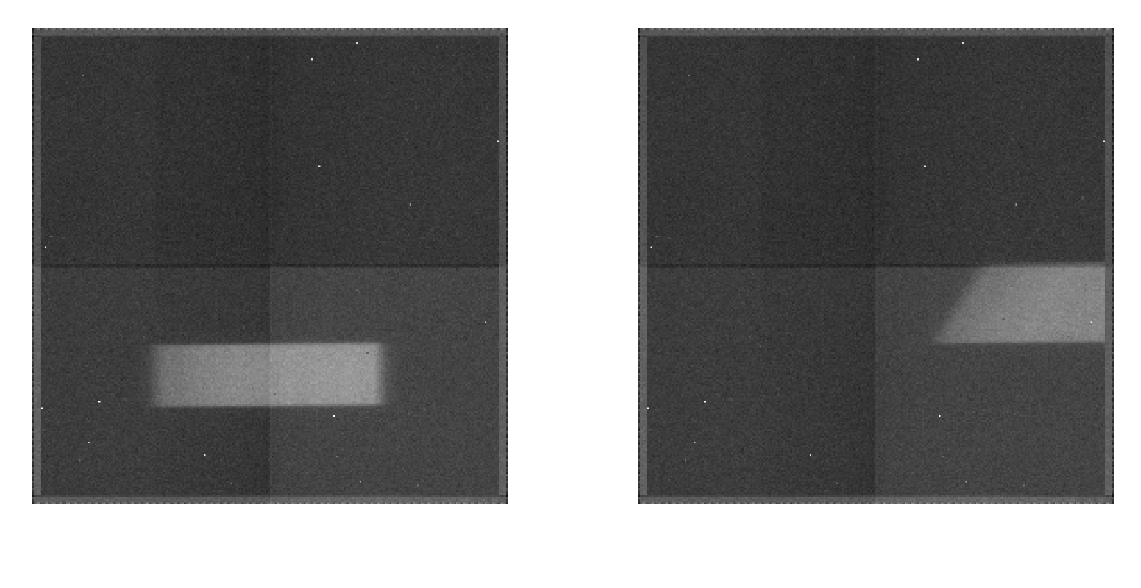}
    \caption{{\textit{Left: } A typical raw \textit{HST} image from Orbits 1-4; \textit{Right: } A typical raw image from Orbits 5-6. The telescope slewed so images from these orbits were discarded.}
    \label{fig:hst_raw_frames}}
\end{figure}

Initial visual inspection of the spectra indicated that the telescope slewed while observing K2-33, resulting in some spectra that are slanted or not visible at all (Figure ~\ref{fig:hst_raw_frames}). There are 6 total orbits. After analyzing the header information, 32 consecutive frames made up of Orbit 5 and Orbit 6 indicated that the guide star acquisition failed and observation was taken only with the gyro guiding. For this reason, these frames were discarded and not used in our analysis. Following the standard procedure \citep{Berta2012}, we excluded the first orbit from our analysis, as it exhibits a much steeper ramp than the subsequent orbits. Of the remaining three \textit{HST} orbits, Orbit 2 provided the out-of-transit baseline, Orbit 3 captured the ingress, and Orbit 4 captured the mid-transit. 

The data set was reduced and calibrated with the publicly available \texttt{Iraclis} pipeline \citep{tsiaras2016new} \footnote{\url{https://github.com/ucl-exoplanets/Iraclis}}. Here, we provide a brief description of their method. We first used the \texttt{\_raw} frames provided by the Hubble Legacy Archive \footnote{\url{https://hla.stsci.edu/hlaview.html}}. The reduction process is performed in the following order: zero-read and bias-level correction, non-linearity correction, dark current subtraction, gain conversion, sky background subtraction, calibration, flat-field correction, and bad pixels and cosmic ray correction. \texttt{Iraclis} then extracted 1D spectra from the reduced data by applying a 128 pixel-wide aperture along the dispersion axis and summing along the cross-dispersion axis for each wavelength. We set the vertical starting and ending position of the extraction box to 5 pixels above and below the spectrum in the spatial direction. There is less than a 0.2 pixel shift in both the x- and y-direction. 

 The \texttt{Iraclis} extraction resulted in a broadband (white) light curve, covering the whole wavelength range in the G141 grism (1.088–1.68$\mu$m) and spectral light curves for each wavelength band. While it is common to adopt a bin width that yields similar SNR in all bins and/or avoid regions of higher stellar variability \citep{Mandell2013}, the partial transit yielded relatively low SNR overall and changes to the bin width did not change any other conclusions about the system. As a result, we adopted 14 equal-size bins, each with a bin width of 0.0423$\mu$m.

\begin{deluxetable} {cccc} 
\tabletypesize{\footnotesize} 
\tablecaption{Observation Log \label{tab:obslog}}
\tablecolumns{7}
\tablewidth{0pt}
\tablehead{
\colhead{Telescope} & 
\colhead{Filter/Grism} &
\colhead{Transit Number } &
\colhead{UT Start Date}}
\startdata
    \ktwo\ Campaign 2 & Kepler & 1-14\tablenotemark{a,b} & 2014 Aug 03\\
    \hline
    \mearth\ North & RG715 & 99\tablenotemark{b,c} & 2016 Feb 16 \\
    & & 104\tablenotemark{b} & 2016 Mar 14 \\ 
    & & 106\tablenotemark{c} & 2016 Mar 25 \\
    & & 111 & 2016 Apr 21 \\ 
    & & 122\tablenotemark{c} & 2016 Jun 20 \\ 
    & & 186 & 2017 Jun 02 \\
    & & 188\tablenotemark{c} & 2017 Jun 13 \\ 
    \hline
    \mearth\ South & RG715 & 104\tablenotemark{b} & 2016 Mar 14 \\
    & & 111 & 2016 Apr 21 \\ 
    & & 186 & 2017 Jun 02 \\ 
    & & 188\tablenotemark{c} & 2017 Jun 13 \\
    & & 191\tablenotemark{c} & 2017 Jun 29 \\ 
    \hline
    \textit{Spitzer}/IRAC & Channel 1 & 148 & 2016 Nov 08 \\
    & & 150 & 2016 Nov 19 \\ 
    & & 188 & 2017 Jun 13 \\
    & & 190 & 2017 Jun 24 \\ 
    & & 217 & 2017 Nov 18 \\
    \hline
    \textit{Spitzer}/IRAC & Channel 2 & 153 & 2016 Dec 06 \\ 
    & & 186 & 2016 Jun 03 \\ 
    & & 187 & 2017 Jun 08 \\ 
    & & 191 & 2017 Jun 30 \\ 
    & & 219 & 2017 Nov 29 \\ 
    \hline
    \hst\ /WFC3 & G141 & 184\tablenotemark{c} & 2017 May 23\\ 
\enddata
\tablenotetext{a}{ There were 14 total consecutive transits taken by \textit{K2}}
\tablenotetext{b}{ Data was already published in \cite{Mann2016b}}
\tablenotetext{c}{ Only a partial transit was observed}
\end{deluxetable}

\section{Updated Stellar Parameters}\label{sec:stellar_params} 
    \citet{Mann2016b} estimated effective temperature (\teff), stellar luminosity ($L_*$), stellar radii ($R_*$), and reddening ($A_V$) for K2-33 using the combination of moderate resolution spectra, magnetic models \citep{Feiden2016}, the transit-fit stellar density (assuming $e=0$), and the distance to the Upper Scorpius OB association \citep{1999AJ....117..354D}. The more precise distance and photometry from \gaia\ Data Release Two \citep[\gaia\ DR2;][]{GaiaDr2} enabled us to improve on these parameters.
    
    We followed the methods in \citet{Mann2016b} for estimating bolometric flux (\fbol\ ) and $A_V$ by simultaneously comparing the flux-calibrated spectra of K2-33 to unreddened young templates and observed photometry of K2-33 from \gaia\ DR2 \citep{Evans2018}, the Two Micron All-Sky Survey \citep[2MASS; ][]{Skrutskie2006}, the Wide-field Infrared Survey Explorer \citep[\textit{WISE}; ][]{allwise}, the Sloan Digital Sky Survey \citep[SDSS; ][]{Abolfathi2018}, and the AAVSO All-Sky Photometric Survey \citep[APASS;][]{Henden2012}. This yielded \fbol\ of $2.40\pm0.18\times10^{-10}$erg cm$^{-2}$ s$^{-1}$ and $A_V=0.89\pm0.13$. Adding the \gaia\ DR2 distance yielded $L_*=0.146\pm0.012L_\odot$, and using the \teff\ from \citet{Mann2016b} and the Stefan-Boltzmann relation gave $R_*=1.017\pm0.057R_\odot$. These are broadly consistent, but more precise than values derived in \citet{Mann2016b}.

    To update $M_*$, we compared all above photometry to the magnetic model grid from \citet{Feiden2016} using a Gaussian prior on age of 10$\pm$3\,Myr and $A_V$ from our spectral fit above. This yielded $M_*=0.571\pm0.054M_\odot$. Other model-based parameters ($R_*$, $L_*$ and \teff) were consistent with our spectral analysis, but we selected the empirical parameters above. Using the $R_*$ above and the model-based mass gave a stellar density ($\rho_*$) of $0.54\pm0.12\rho_\odot$. A summary of our adopted stellar parameters used in our transit analysis are listed in Table~\ref{tab:stellar_param}
   
\begin{deluxetable} {l|cc}
\tablecaption{Updated Stellar Parameters for K2-33
\label{tab:stellar_param}}
\tablecolumns{3}
\tablewidth{0pt}
\tablehead{
\colhead{Parameters} & 
\colhead{Value} &
\colhead{Source}}
\startdata
    $A_{v}$ & 0.89 $\pm$ 0.13 & This Paper \\
    \fbol\ ($\times 10^{-10} $\,erg\,cm$^{-2}$\,s$^{-1}$) & 2.40$\pm$ 0.18 & This Paper\\ 
    $R_*$ $(R_\odot)$ & 1.017$\pm$0.057 & This Paper \\
    $M_*$ $(M_\odot)$ &  0.571$\pm$0.054 & This Paper \\
    $\rho_*$ $(\rho_\odot)$ &  0.54$\pm$0.12 & This Paper \\
    $L_*$ $(L_\odot)$  & 0.146$\pm$0.012 & This Paper \\
    $\teff$ (K) & 3540$\pm$70 & \cite{Mann2016b}  \\
\enddata
\end{deluxetable}
    
    We estimated limb-darkening parameters for all observations using the above stellar parameters and the \texttt{LDTK} toolkit \citep{2015MNRAS.453.3821P}\footnote{v1.5, which added support for limb-darkening parameters past 5\um}, which uses PHOENIX models \citep{2013A&A...553A...6H} and propagated uncertainties in stellar parameters onto the output limb-darkening values. Different models give different limb-darkening parameters at the level 0.04--0.08, so we inflated these errors by those values to account for differences between input stellar models. For MEarth, {\it K2}, and both \spitzer\ bands, we opted for a two-parameter limb-darkening value ($g_{1}$, $g_{2}$). For {\it HST}, we used the non-linear (4-parameter) limb-darkening formula detailed in \citet{2000A&A...363.1081C} ($a_{1}$, $a_{2}$, $a_{3}$, $a_{4}$). In each case, we used the relevant filter profile from the instrument documentation. To simulate the {\it HST} spectral bands, we multiplied the G141 filter profile with top-hat profiles corresponding to the limits of each spectral band. The resulting limb-darkening estimates were fed directly into \texttt{MISTTBORN} (\ktwo, \mearth, and \spitzer) or \texttt{Iraclis} (\hst), as described in the next section. A summary of our adopted limb darkening coefficients used in our transit analysis are listed in Table~\ref{tab:limb_darkening_priors}. 

\section{Transit Fitting}\label{sec:transit_fitting} 

\subsection{\ktwo\ and MEarth}\label{sec:mearthktwo}

We fit both the \ktwo\ and MEarth data individually using the \texttt{MISTTBORN} (MCMC Interface for Synthesis of Transits, Tomography, Binaries, and Others of a Relevant Nature) fitting code\footnote{\url{https://github.com/captain-exoplanet/misttborn}} described in \citet{Mann2016a} and in more detail in \citet{MISTTBORN}. \texttt{MISTTBORN} uses \texttt{BATMAN} \citep{Kreidberg2015} to generate model transit curves and uses \texttt{emcee} \citep{Foreman-Mackey2013} to explore the transit parameter with a MCMC algorithm.

With \texttt{MISTTBORN}, we fit four parameters directly relating to the transiting planet: time of periastron ($T_0$), orbital period of the planet ($P$), planet-to-star radius ratio ($R_p/R_\star$), and impact parameter ($b$). For the \ktwo\ data, all of these parameters evolved under uniform priors. For the \mearth\ data, we placed a weak Gaussian prior on $T_0$ and $P$ ($\pm$0.01\,days) to prevent the solution from wandering away from the data; other parameters evolved under uniform priors. An additional parameter, stellar density ($\rho_\star$), allowed us to replace the more common transit duration and impose a Gaussian prior from our stellar parameters in Section~\ref{sec:stellar_params}. For each of the two wavelengths, we fit two linear and quadratic limb-darkening coefficients ($q_1$, $q_2$) following the triangular sampling prescription of \citet{Kipping2013}. Both evolved under Gaussian priors derived from \texttt{LDTK} (see Section~\ref{sec:stellar_params}). We locked the eccentricity ($e$) to zero because gas drag and gravitational interactions are expected to dampen out eccentricities and inclinations of extremely young planets like K2-33b \citep{2004ApJ...602..388T}. In the case that eccentricity is non-zero, this does change the overall depth, but it does not change the relative depths because all impact parameters will move together.

An important difference between our analysis here and that of \citet{Mann2016a} was that we fit the stellar variability simultaneously with the transit parameters. This change was driven, in part, because of cases where our removal of stellar variability would impact the transit, yielding a systematically smaller transit depth than when fitting simultaneously. For example, the transit depths determined by the two discovery papers differed slightly despite using overlapping data \cite[ 0.23 vs. 0.19\%;][]{David2016, Mann2016b}. 

To this end, we used the \texttt{celerite} package \citep{celerite} for Gaussian Process fitting which has now been incorporated into \texttt{MISTTBORN}. We first tested the ``SHOM'' kernel, i.e., a mixture of two stochastically driven damped simple harmonic oscillators (SHO) with periods $P$ and $0.5P$. However, we found the parameters of the second SHO (at half the period) was poorly constrained, which is likely because the primary rotation signal dominates \citep{nicholson2022quasi}. Instead, we adopted a single SHO fit, which added three free parameters: the log of period ($\ln(P_{GP})$), the log of the variability amplitude ($\ln{A}$), and the decay timescale ($\ln{Q}$). 

When fitting the \ktwo\ data, we evolved all GP parameters under uniform priors. For the \mearth\ data, there is insufficient out-of-transit data to constrain the GP, so we applied a Gaussian prior on $\ln(P_{GP})$ and $\ln{A}$ based on the output from the \ktwo\ fit. Since the two datasets are at slightly different wavelengths (0.6\um\ vs 0.8\um) the spot-variability amplitude might not be the same (and hence so might be the GP amplitude). However, factor-of-two changes to the amplitude prior in either direction did not change the resulting transit depths, most likely because the GP is able to adjust to each transit between the long data gaps between MEarth observations. 

For both datasets, we ran \texttt{MISTTBORN} with 100 walkers for 250,000 steps after a burn-in of 50,000 steps. A comparison to the autocorrelation time suggested this was more than sufficient for convergence \citep{2010CAMCS...5...65G}. We summarize the \texttt{MISTTBORN} output in Table~\ref{tab:misttborn_fits}. The fit to the \ktwo\ data is shown in Figure~\ref{fig:k2_with_gp}.

\begin{table*} 
\centering
\caption{Parameters of K2-33 for Different Data Sets}\label{tab:misttborn_fits} 
\begin{tabular}{llcccc} 
\hline 
\hline 
Description & Parameter & \ktwo & \mearth  & \spitzer\ [3.6] & \spitzer\ [4.5]   \\ 
\hline 

First Transit Mid-point & $T_0$ (KBJD\footnote{KBJD = BJD-2454833}) & $2065.6921^{+0.0028}_{-0.0027}$ &  $2065.6923 \pm 0.0045$ &  $2065.6982^{+0.0046}_{-0.0048}$ & $2065.6954^{+0.0079}_{-0.0081}$  \\ 
Orbital Period & $P$ (days) & $5.42503 \pm 0.00034$ & $5.424832^{+2.6\times10^{-5}}_{-2.5\times10^{-5}}$ & $5.424851^{+2.6\times10^{-5}}_{-2.7\times10^{-5}}$ & $5.424862^{+4.2\times10^{-5}}_{-4.1\times10^{-5}}$\\ 
Planet-Star Radius Ratio & $R_P/R_{\star}$ & $0.04735^{+0.00099}_{-0.00096}$ & $0.0489^{+0.0019}_{-0.0020}$ & $0.03545 \pm 0.00086$ & $0.03746^{+0.00083}_{-0.00079}$ \\
Impact Parameter & $b$ & $0.19^{+0.18}_{-0.13}$  & $0.27^{+0.2}_{-0.18}$&  $0.24^{+0.21}_{-0.17}$ & $0.22^{+0.2}_{-0.15}$ \\ 
Stellar Density & $\rho_{\star}$ ($\rho_{\odot}$) & $0.464^{+0.042}_{-0.066}$ & $0.527^{+0.062}_{-0.098}$  & $0.483^{+0.04}_{-0.1}$ & $0.476^{+0.038}_{-0.088}$ \\ 
Limb-darkening coefficient & $q_{1}$ & $0.48^{+0.083}_{-0.082}$ & $0.401^{+0.079}_{-0.078}$ & $0.072^{+0.033}_{-0.031}$ & $0.06^{+0.027}_{-0.024}$ \\ 
Limb-darkening coefficient & $q_{2}$ & $0.351 \pm 0.046$ & $0.286 \pm 0.048$ & $0.209^{+0.033}_{-0.034}$ & $0.204 \pm 0.034$ \\ 
Log Period & $\ln{P}$ & $1.838 \pm 0.015$ & $1.836 \pm 0.015$ &  $2.6^{+2.57}_{-0.83}$ & $0.99^{+0.68}_{-0.4}$ \\ 
Log Variability Amplitude & $\ln{A}$ & $-8.96^{+0.26}_{-0.22}$ & $-10.14^{+0.44}_{-0.39}$& $-13.84^{+1.04}_{-0.86}$ & $-13.93^{+1.12}_{-0.92}$\\ 
Log Decay Timescale & $\ln{Q}$ & $0.926^{+0.13}_{-0.087}$ & $0.048^{+0.083}_{-0.036}$&  $6.2^{+2.9}_{-3.0}$ & $4.5^{+3.3}_{-2.9}$\\ 
\hline 
\end{tabular} 
\end{table*}

\subsection{\spitzer}\label{sec:transit_fit_spitzer}
\textit{Spitzer}’s large intra-pixel sensitivity and pointing jitter can cause the measured flux of a source to vary up to 8\%, depending on where it falls on a pixel \citep{Ingalls2012}. However, years of high-precision observations with \spitzer\ have provided a number of methods to correct for model variations in the photometric response \citep{Ingalls2016}. We tested multiple methods, which includes using a high-resolution pixel variation gain map \citep[PMAP;][]{Ingalls2012} and nearest neighbors \citep[NNBR;][]{Lewis2013}, but found the most consistent results using the BiLinearly-Interpolated Subpixel Sensitivity \citep[BLISS;][]{StevensonTransit2012} mapping technique, which we briefly summarize below. We note that changing to NNBR did not change any of the conclusions, only the overall fit quality. 

The BLISS Mapping Technique is provided as an optional part of the \texttt{POET} pipeline. It corrects for both the position-dependent (intra-pixel effect) and time-dependent (ramp effect) \spitzer\ systematics. The parameters in the BLISS mapping include the grid size of the sub-pixel sensitivity map, the astrophysical light curve model (e.g., transit, secondary eclipse), and the ramp model.

For the Channel 1 data, to avoid overfitting, we followed \texttt{POET}'s recommendation of choosing a grid size in which a nearest-neighbor interpolation would not outperform a BLISS interpolation \citep{StevensonTransit2012}. We tested various grid sizes and selected 0.007 pixels. For the Channel 2 data, we used the most recent fixed intra-pixel sensitivity map from \cite{MayIntroducing2020}.

The transit was modeled using the \cite{MandelAgol2002} transit model and several different ramp parameterizations were tested: linear, quadratic, rising exponential, falling exponential, quartic-log + quadratic polynomial, log + linear ramp, and a no-ramp model. For further information about each individual ramp, see \cite{StevensonTransit2012}. For each light curve, we determine the best ramp model based on three metrics: 1) the difference in predicted t0 value compared to the expected t0, 2) the difference in the predicted transit duration compared to the expected transit duration, and 3) the overall minimal red noise levels in the fit residual, assessed by considering the root-mean-squared (RMS) binned residuals as a function of different bin sizes with the theoretical uncorrelated white noise. A small $T_0$ difference, small duration difference, and low RMS were all favored.

The time-dependent component of the transit model consisted of the mid-transit time ($T_{0}$), planet-to-star radius ratio ($R_{p}/R_{s}$), orbital inclination ($\cos{i}$), semi-major axis ratio ($a/R_{*}$), and system flux ($\mu Jy$), as well as, parameters associated with the ramps (e.g., ramp phase, ramp amplitude, ramp constant offset, etc.). These parameters were explored with an MCMC process, using 4 walkers with 200,000 steps and a burn-in region of 30,000 steps. The period was fixed to 5.4248 days, based on our analysis of the \ktwo, \mearth, and an initial reduction of the \spitzer\ data (which provided a sub-second precision period). As recommended by \cite{MayIntroducing2020}, temporal binning was not used for Channel 2. The resulting \texttt{POET} fits for each light curve are summarized in Table~\ref{tab:poet_fits}. 

\begin{deluxetable*}{l|ccr|ccccr}
\tabletypesize{\scriptsize}
\label{tab:poet_fits}
\tablecaption{Best Fit Parameters for {\spitzer\  Data Using \texttt{POET}} 
}
\tablehead{
\colhead{AOR} &
\colhead{Transit Num} &
\colhead{BIC Value} &
\colhead{Best Ramp} & 
\colhead{T$_{0}$ [BJD]} & 
\colhead{cos\textit{i}} & 
\colhead{a/R$_{*}$} & 
\colhead{System Flux [$\mu Jy$]} & 
\colhead{R$_{p}$/R$_{*}$} 
} 
\startdata
\multicolumn{9}{c}{3.6 $\mu$m}\\
\hline
60655360 & 148 & 10773 & quadratic & 2457701.57 & 0.07 & 8.46 & 31325 & 0.037 $\pm$ 0.004 \\
60658432 & 150 & 10765 & quadratic & 2457712.42 & 0.07 & 8.79 & 31172 & 0.063 $\pm$ 0.002 \\
60661504 & 188 & 14433 & quartic-log + quartic polynomial &  2457918.57 & 0.00 & 10.52 & 30291 & 0.035 $\pm$ 0.003 \\
60664576 & 190 & 14396 & linear  & 2457929.41 & 0.03 & 10.00 & 29864 & 0.036 $\pm$ 0.002 \\ 
60664576 & 217 & 14407 & quartic & 2458075.89 & 0.07 & 8.63  & 30618 & 0.035 $\pm$ 0.002 \\ 
\hline
\multicolumn{9}{c}{4.5 $\mu$m}\\
\hline
60656128 & 153 & 10754 & quadratic & 2457728.70 &  0.00 & 10.50 & 19369 & 0.050 $\pm$ 0.003 \\
60659200 & 186 & 14406 & log & 2457907.72 & 0.07 & 8.61 & 19550 &  0.037 $\pm$ 0.004 \\ 
60662272 & 187 & 14390 & quadratic & 2457913.15 & 0.00 & 10.27 & 19662 & 0.038 $\pm$ 0.003 \\ 
60665344 & 191 & 14390 & rising exponential & 2457934.85 & 0.07 & 8.54 & 19677 &  0.037 $\pm$ 0.002\\ 
60668416 & 219 & 14388 & rising exponential & 2458086.74 & 0.07 & 8.80 & 19664  & 0.041 $\pm$ 0.001 \\
\hline
\enddata
\end{deluxetable*}

\begin{figure*} [hpt]
    \centering
    \includegraphics[trim=0cm 5cm 0cm 5cm, clip=True, width=500pt]{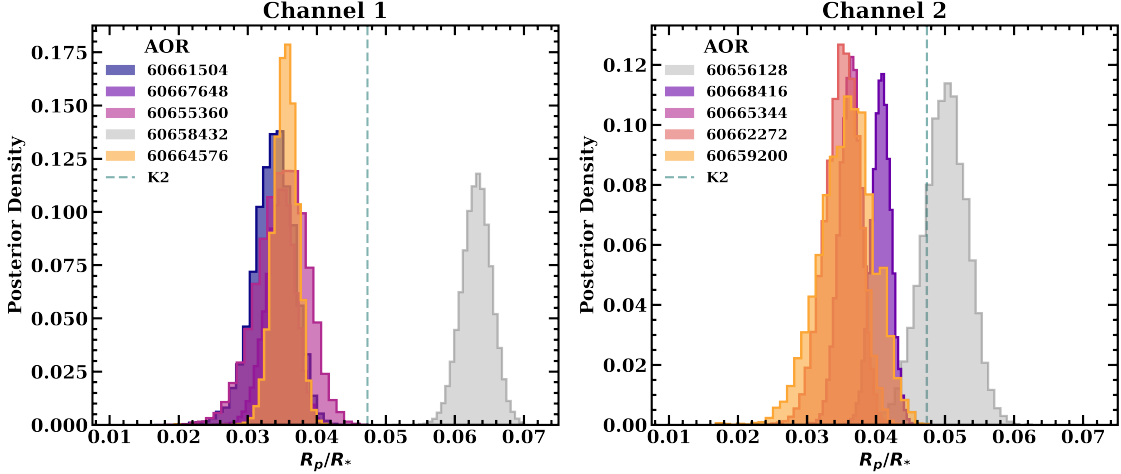}
    \caption{Planet-to-star radius ratio ($R_{p}/R_{*}$) posteriors from the \texttt{POET} MCMC fits for \spitzer\  Channel 1 (\textit{left}) and Channel 2 (\textit{right}) with a bin size of 30. Each color represents a different transit (AOR), where gray is used to highlight the outlier transit in each channel (AOR 60658432 in Channel 1 had a flare and AOR 60656128 in Channel 2 had poor pointing). The green dashed vertical line is the planet-to-star radius ratio for \textit{K2}, which does not overlap with any of the \textit{Spitzer} datapoints, except for the outlier transit in Channel 2.
    \label{fig:rprs_spitzer}}
\end{figure*}

\begin{figure}[ht]
    \centering
    \includegraphics[width=0.49\textwidth,height=.49\textwidth]{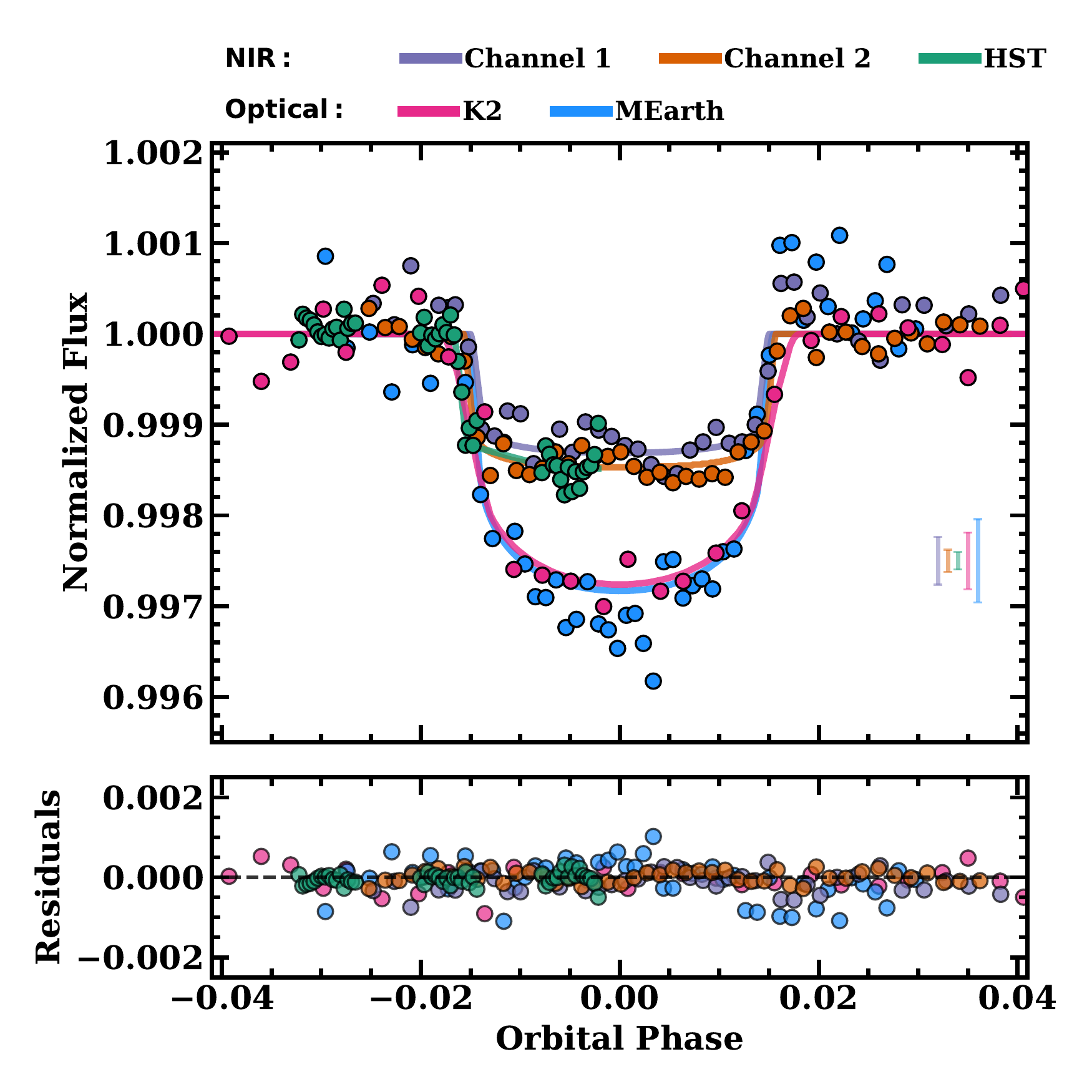}
    \caption{\textit{Top: } Stacked phase-folded light curve of K2-33 observed in \textit{K2} (pink), MEarth (blue), \textit{Spitzer}/Channel 1 (purple), \textit{Spitzer}/Channel 2 (orange), and \textit{HST} (green). The \textit{HST} data here is the white light curve. Data points correspond to the light curve binned in phase using a bin size of 10, 200, 1500, and 1500 for \textit{K2}, MEarth, Channel 1, and Channel 2, respectively. Solid line corresponds to the best-fit (highest likelihood) model from our MCMC fit. Typical error bars are derived from scatter in the out-of-transit data points. The transit depth for the optical wavelengths is $\sim$ 2 times deeper than the transit depth for the near-infrared wavelengths. \textit{Bottom: } Residuals using the binned points. 
    \label{fig:lc_comparison}}
\end{figure}

A drawback of the \texttt{POET} results is that we could not fit all transits simultaneously.  This is particularly important since several of the \spitzer\ transits yield discrepant depths (Figure ~\ref{fig:rprs_spitzer}). To this end, we re-fit the output light curves from \texttt{POET} with \texttt{MISTTBORN}, as described in Section~\ref{sec:mearthktwo}. We fit the two channels separately, including all but one transit in each band. The excluded observations were AOR 60658432 from Channel 1 (which had a strong flare) and AOR 60656128 from Channel 2 (which had extremely poor pointing). These two transits also yielded outlier transit depths from \texttt{POET} (see Figure~\ref{fig:rprs_spitzer}). 

Unlike with \texttt{POET}, \texttt{MISTTBORN} cannot simultaneously fit for instrumental effects with the transit, so we attempt to account for that by including the GP. We tried several GP kernels, including Matern-3/2 and a simple white noise term, but found the SHO we used for stellar variability worked as well or better. Since the periodic signal in this case is instrumental, not stellar, we allow all GP parameters to evolve under uniform priors (instead of forcing it to match the rotation period, for example). As above, the only transit parameters we place priors on were limb-darkening and stellar density.

The resulting \texttt{MISTTBORN} fits are summarized in Table~\ref{tab:misttborn_fits}. Other than depth, the transit-specific parameters were in broad agreement with those from \ktwo\ and \mearth. Excluding the two problematic transits, the transit depths from our \texttt{MISTTBORN} fits are in excellent agreement with those from \texttt{POET}. The \texttt{MISTTBORN} fits gave similar or smaller uncertainties than individual \texttt{POET} fits, but still larger than expected from a simple weighted mean of the individual \texttt{POET} fits. We consider the \texttt{MISTTBORN} fits to be more realistic. 

Importantly, both the \texttt{MISTTBORN} and \texttt{POET} fits to the data yielded a transit depth $\simeq$half that from the optical data. This difference was not explainable through the fitting, as it is clear even in the extracted light curves (Figure~\ref{fig:lc_comparison}). Further, as we will show in Section~\ref{sec:HST}, the effect is seen in the \hst\ data as well.

\subsection{\hst}\label{sec:HST}

\subsubsection{White Light Curve Fitting}

In order to fit the extracted light curves, we must first correct the time-dependent systematics (ramps) introduced by the WFC: 1) a long term ramp that occurs throughout the visit and typically has a linear behavior and 2) a short term ramp that occurs during each orbit and typically has an exponential behavior \citep{tsiaras2016new}.

\begin{figure}[ht!]
    \includegraphics[width=0.49\textwidth,height=.49\textwidth]{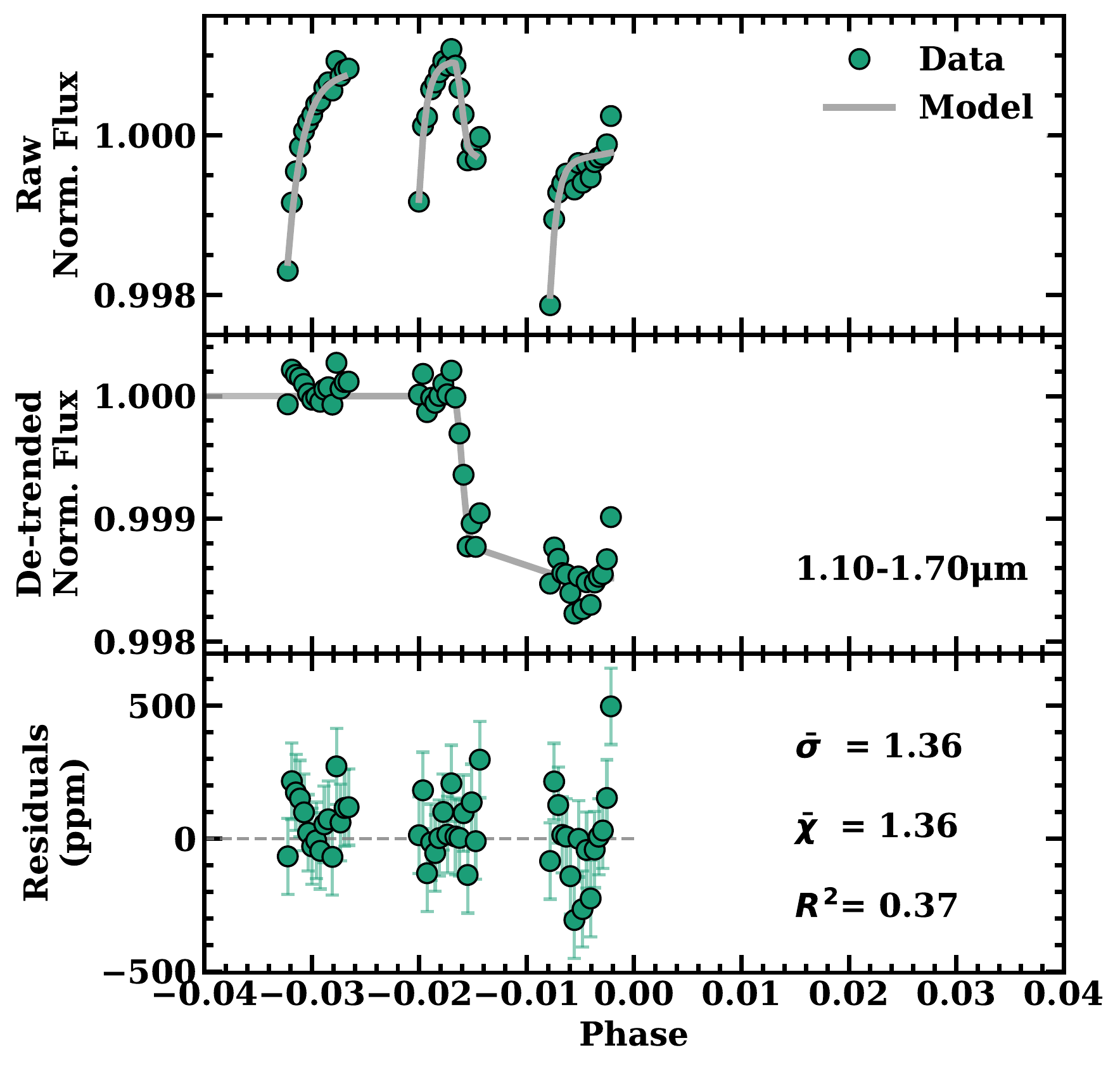}
    \caption{White light curve of K2-33 observed with \textit{HST/WFC3}. The top panel shows the raw normalized data (green circles) overplotted with the systematic model (gray line). The middle panel shows the best fit model (gray line) from our MCMC fit overplotted with the systematic-corrected data (green circles). The bottom panel shows residuals from the light curve.\label{fig:hst_white_lc}} 
\end{figure}

The light curve is fitted using a transit model from \texttt{PyLightcurve} \footnote{\href{https://github.com/ucl-exoplanets/pylightcurve}{https://github.com/ucl-exoplanets/pylightcurve}}, and with an instrumental systematics function $R(t)$ \citep{tsiaras2016new} : 

\begin{equation}
    R_{w}(t) = n_{w} (1-r_{a1}(t-T_{0})) (1-r_{b1} e^{-r_{b2}(t-t{_0})})
\end{equation}

where $n_{w}$ is a normalization factor, $t$ is time of the data, $T_{0}$ is the model mid-transit time, $t_{0}$ is the starting time of each \textit{HST} orbit, $r_{a1}$ is the slope of the linear systematic trend, $r_{b1}$ and $r_{b2}$ are the coefficients of the exponential systematic trend. Since the first orbit ramp (Orbit 2) is steeper compared to the ramps of the subsequent orbits (Figure~\ref{fig:hst_white_lc}), a different set of the short-term exponential coefficients ($for_{b1}$ and $for_{b2}$) were used to fit this ramp. Removing these coefficients will increase the root-mean-square (RMS) of the residuals.

Due to only having half a transit and missing both the egress and out-of-transit baseline, we limit our ability to constrain the orbital parameters. Therefore, we locked the inclination($i$), the semi-major axis-to-star radius ratio ($a/R_{*}$), the period ($P$), the argument of periastron ($\omega$), and the eccentricity ($e$) to the values from a combined fit of data from  \ktwo, \mearth, and  \textit{Spitzer}. Limb-darkening coefficients ($a$) were fixed to the model-derived values (Table ~\ref{tab:limb_darkening_priors}). 

The only free parameters were coefficients for \textit{Hubble} systematics, the normalization factor ($n_{w}$), the planet-to-star radius ratio ($R_{p}/R_{*}$), and the mid-transit ($T_{0}$). These parameters were explored with an MCMC process, using 200 walkers with 40,000,000 steps, and a burn-in region of 6,000,000 steps. This was sufficient for convergence based on the autocorrelation time. The final $R_{p}/R_{*}$ result for the broadband (white) light curve can be found in Table ~\ref{tab:limb_darkening_priors}. 

The fit white light curve is shown in Figure~\ref{fig:hst_white_lc}. As can be seen in the bottom panel, the residuals are within 300 ppm for both Orbit 2 and Orbit 3, but not for Orbit 4, due to the remaining uncorrected systematics. 

The autocorrelation of the residuals was 0.37. This value is slightly above the threshold used by \cite{tsiaras2018population}, who defined a fit to be successful if the autocorrelation was $<$ 0.3. In that work, they applied \texttt{Iraclis} to 30 gaseous planets taken by \textit{HST/WFC3}. However, their study was focused on full transits and targets significantly less variable than K2-33. Our standard deviation of the residuals relative to the photon noise ($\bar{\sigma}$) is 1.36 and this is within the reasonable limits from \cite{tsiaras2018population} (1.10 $ < \bar{\sigma} <$ 2.75). We considered this fit to be reasonable, but given the partial transit loss, more data would be required to further investigate the fit quality.

\subsubsection{Spectral Light Curve Fitting}

To extract the planetary spectrum from the spectral light curves, we must first correct for the systematics. \texttt{Iraclis} fits the spectra light curve using the 
\textit{divide white method}, first introduced by \citet{Kreidberg2014}, in which a transit mode is multiplied with an instrumental systematics function $R_{\lambda} (t)$: 
\begin{equation}
    R_{\lambda}(t) = n_{\lambda} [1-r_{a, \lambda}(t-T_{0})]\frac{LC_{W}}{M_{W}},
\end{equation}
where $n_{\lambda}$ is the normalization factor for each wavelength bin, $r_{a1}$ is the linear long term ramp, $LC_{W}$ is the white light curve, and $M_{W}$ is the model of the white light curve. 

\begin{figure*} [htp!]
    \centering
    \includegraphics[width=0.8\textwidth,height=.8\textwidth]{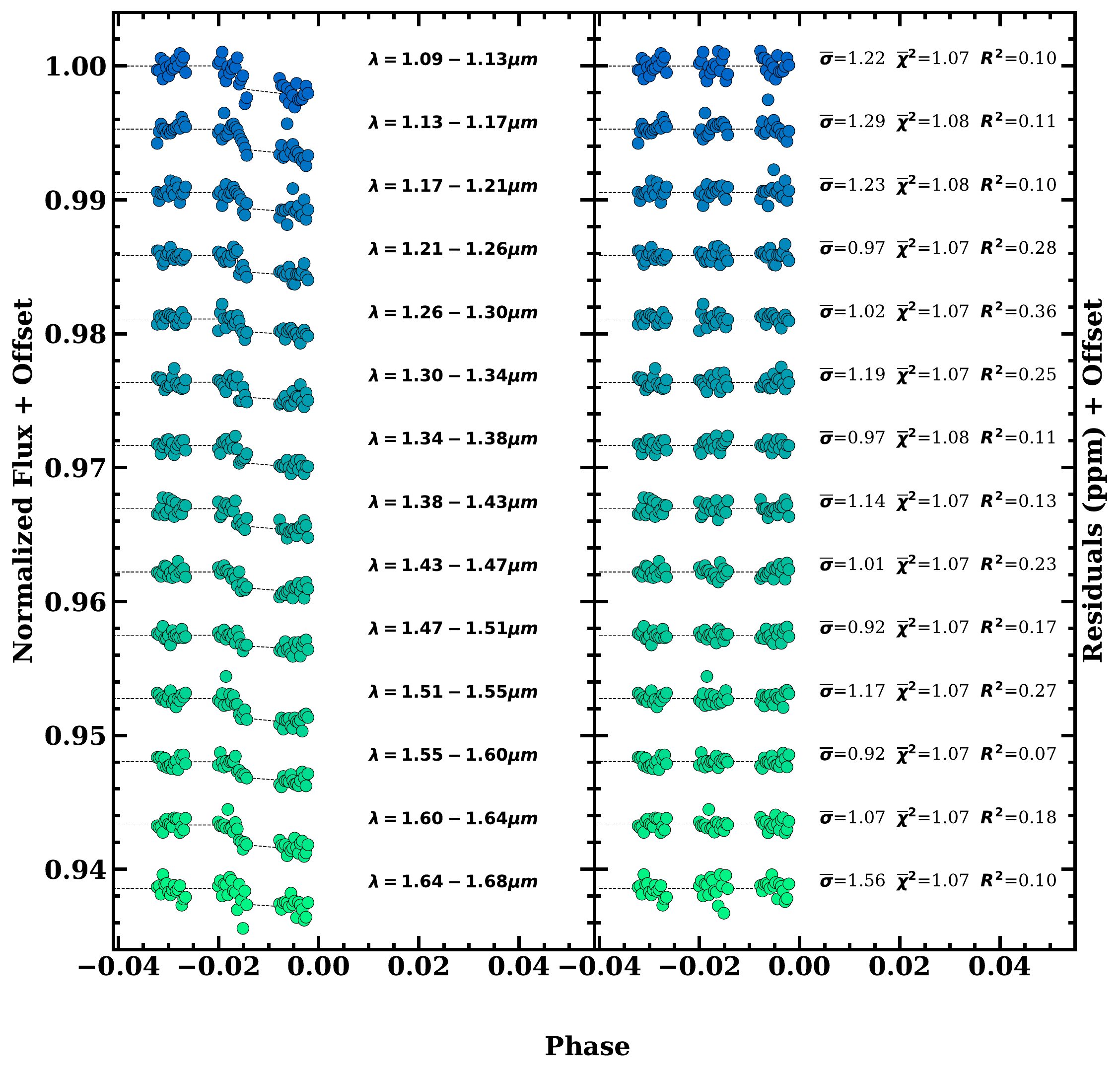}
    \caption{\textit{Left}: Spectral light curve annotated with the mean wavelength for each bins. \textit{Right}: Residuals of the fit, along with the standard deviation relative to the photon noise ($\bar{\sigma}$), the reduced chi-square ($\bar{\chi}^2$), and the square of the correlation ($R^{2}$) \label{fig:hst_spectral_lc}}
\end{figure*}

The transit model was fixed to the same orbital parameters as the white light curve. We did not fit for the limb-darkening coefficients. The only free parameters are the linear long term ramp, the normalization factor, and the planet-to-star radius ratio ($R_{p}/R_{*}$). The parameter space was explored with an MCMC process using 100 iteration walkers, 10,000,000 steps and a burn-in region of 2,000,000 steps. The final $R_{p}/R_{*}$ results for each spectral bin and their uncertainties can be found in Table~\ref{tab:limb_darkening_priors}. All the spectral light curves for each wavelength bins are plotted in Figure ~\ref{fig:hst_spectral_lc}. The standard deviation of the residuals is on average 1.12 times the expected photon noise ($\bar{\sigma}$), and the residuals autocorrelation ($R^{2}$) is 0.18. Both these values fall within the range of the spectral light curve fits from \cite{tsiaras2018population} ($\bar{\sigma} < 1.17 $ and $R^{2} < 0.20$).

\begin{deluxetable*} {lcccccccc}
\tabletypesize{\scriptsize}
\tablecaption{Limb-darkening Coefficients For Each Data Set \label{tab:limb_darkening_priors}}
\tablecolumns{7}
\tablewidth{20pt}
\tablehead{
\colhead{Telescope} &
\colhead{$\lambda$ range ($\micron$)} 
& \colhead{$g_{1}$} &
\colhead{$g_{2}$} &
\colhead{} &  
\colhead{} & 
}
\startdata
\textit{K2} & 0.42--0.90 &  $0.479\pm0.050$ & $0.211\pm0.036$ \\
MEarth & 0.69--1.00 & $0.317\pm0.048$ & $0.274\pm0.035$\\ 
\textit{Spitzer} Channel 1 & 3.13--3.96 & $0.112\pm0.015$ & $0.155\pm0.019$\\
\textit{Spitzer} Channel 2 & 3.92--5.06 & $0.085\pm0.011$ & $0.124\pm0.018$ \\
\hline
Telescope & $\lambda$ range ($\micron$) & $a_{1}$ & $a_{2}$ & $a_{3}$ & $a_{4}$ & {$R_{p}/R_{*}$}\\ 
\hline
\textit{HST} & 1.0880--1.1303 & -0.133292 & 1.848572 & -1.917848 & 0.665771 & $0.043544^{+0.004269}_{-0.004879}$ \\ 
& 1.1303--1.1726 & -0.121481 & 1.783907 & -1.858905 & 0.646929 & $0.040560^{+0.004145}_{-0.004737}$ \\ 
& 1.1726--1.2149 & -0.115123 & 1.742967 & -1.820934 & 0.634099 & $0.035911^{+0.004172}_{-0.004768}$ \\ 
& 1.2149--1.2571 & -0.107202 & 1.694365 & -1.767156 & 0.614537 & $0.036135^{+0.00303}_{-0.00303}$ \\ 
& 1.2571--1.2994 & -0.100966 & 1.657434 & -1.729879 & 0.601442 & $0.032079^{+0.003439}_{-0.003439}$ \\ 
& 1.2994--1.3417 & -0.106114 & 1.672739 & -1.746672 & 0.606971 & $0.034728^{+0.003642}_{-0.003642}$ \\ 
& 1.3417--1.3840 & -0.137048 & 1.853932 & -1.936428 & 0.671591 & $0.037669^{+0.002757}_{-0.00315}$ \\ 
& 1.3840--1.4263 & -0.150512 & 1.907895 & -1.981589 & 0.684045 & $0.037647^{+0.003729}_{-0.003262}$ \\ 
& 1.4263--1.4686 & -0.158651 & 1.977145 & -2.061429 & 0.710897 & $0.035890^{+0.003073}_{-0.003512}$ \\ 
& 1.4686--1.5109 & -0.154233 & 2.000803 & -2.104788 & 0.727856 & $0.029557^{+0.003642}_{-0.003187}$ \\ 
& 1.5109--1.5531 & -0.149891 & 2.013220 & -2.127071 & 0.737913 & $0.040000^{+0.00318}_{-0.00318}$ \\ 
& 1.5531--1.5954 & -0.123405 & 1.967341 & -2.120444 & 0.742160 & $0.035862^{+0.003098}_{-0.002711}$ \\ 
& 1.5954--1.6377 & -0.092558 & 1.879567 & -2.055896 & 0.726144 & $0.039918^{+0.00286}_{-0.00286}$ \\ 
& 1.6377--1.6800 & -0.057187 & 1.743134 & -1.937097 & 0.689630 & $0.035785^{+0.005314}_{-0.005314}$ \\ 
\hline
\textit{HST} White Light Curve & 1.0880--1.6800 & -0.127289 & 1.860330 & -1.965358 & 0.685232 & $0.036702^{+0.001698}_{-0.001485}$ \\
\enddata
\tablecomments{Limb darkening values for all dataset were calculated using the \texttt{LDTK} toolkit \citep{2015MNRAS.453.3821P}. For the \ktwo\ , \mearth\ , and \spitzer\ data set, limb darkening priors are provided as the traditional linear and quadratic terms, but were fit using triangular sampling terms.}
\end{deluxetable*} 

Overall, the \hst\ data was too imprecise to confidently detect features, particularly given the loss of egress. However, even the partial transits confirm the large difference between the optical and NIR transit depths; \hst\ and \spitzer\ both yielded transit depths almost half as deep as the depths from \ktwo\ and MEarth. This rules out that the depth difference is due to unconstrained systematics in the datasets. We explored two possible reasons for this difference: spots on the stellar surface (Section~\ref{sec:transit_spots}) and hazes in the planet's atmosphere (Section~\ref{sec:hazes}). 

\section{The effect of surface inhomogeneities on the transmission spectrum}\label{sec:transit_spots}

Unocculted spots on the stellar surface can make a transit appear deeper. If the transit chord has fewer spots, the planet will block a statistically warmer (brighter) region of the star. The effect is reversed for unocculted plages. Since surface inhomogeneities vary in intensity with wavelength, the effect is wavelength-dependent and can mimic or complicate transmission spectroscopy of the planet \citep[e.g., ][]{ 2015ApJ...814...66K, rackham2017access}. Similarly, occulted spots or plages can make the transit shallower or deeper, respectively. We show below that the large difference between the optical and NIR transit depths could be explained if more than half ($>$60\%) of the stellar surface is covered in cool spots and the transit chord is clear, or if the transit cord is heavily populated by plages while the rest of the star is clear. 

\subsection{Planet crossing over spots and plages}
If the planet were crossing a significant number of spots or plages during transit, we would expect to see two effects. First, we would see distortions in the light curve during transit \citep[e.g., as seen for HAT-P-11 in ][]{2017ApJ...846...99M}. While this would be a challenge to see with long-cadence (cadence = 1765.5 s) data from {\it K2}, it should be readily visible in the high-cadence (cadence = 60 s) MEarth data. The second effect should be variations in the transit depth between transits due to differences between the stellar and orbital periods; each transit crosses a slightly different region of the star, changing the properties of the occulted surface. Changes in the underlying spot pattern are also visible in the the {\it K2} light curve (Figure~\ref{fig:k2_with_gp}), likely due to differential rotation \citep[as noted in ][]{David2016b}. However, individual transit depths are consistent with each other within measurement errors, the scatter in the light curve during transit is the same as outside of the transit, and there are no clear morphological changes in the \mearth\ or \ktwo\ data during transit. Planet-crossing plages would also need to be long-lived to explain both the {\it K2} and \mearth\ data, which spans 2.7 years and has more than 150 full stellar rotations. 

Due to the reasons above, we assumed that the transit chord is pristine for the rest of our analysis and focused on the effects of unocculted spots on the rest of the star. We note that consistency between transit depths with time disfavors significant unocculted spots, since it is unlikely that any given transit would have the same spot coverage fraction. However, this is easier to get around with polar or symmetric spot distributions and small variations in the spot coverage fraction may get lost in the noise.

\subsection{Unocculted spots}

\subsubsection{What kind of spot coverage can reproduce the observed transit depths?}\label{sec:lightsource}

Faculae have shown to have less of an effect in the stellar variability in the \kepler\ light curves, justifying our reasons for not including them in our analysis \citep{johnson2021forward}. To estimate what kind of spots are required to reproduce the observed transit depths, we followed a modified procedure from \citet{rackham2017access}, \citet{Thao2020}, and \citet{2021arXiv210510487L}. To summarize, we assumed the surface can be described by a simple two-temperature model of three parameters: the surface temperature (\tsurf), the spot temperature (\tspot), and the fraction of the star with \teff$=$\tspot\ ($f_S$). 

As explained by \citet{2021arXiv210510487L}, $f_S$ is actually time variable for these two reasons: 1) it represents the surface seen from Earth, therefore $f_S$ changes as the star rotates, and 2) the overall surface spot pattern can change with time. We ignore this effect for now. As we will show in Section~\ref{sec:lightcurve}, the overall spot coverage needs to be relatively stable over the rotational period, or the light curve would show more significant variability in the \ktwo\ data. 

We built our synthetic star using surface brightness estimates from PHOENIX/BT-SETTL \footnote{\url{https://phoenix.ens-lyon.fr/Grids/BT-Settl/CIFIST2011_2015/}} atmosphere models \citep{Allard2013,Allard2012}. For both the spots and surface, we assumed Solar abundances and adopted simple linear interpolation between grid points in \teff. Surface gravity ($\log(g)$) has negligible effects on the spectrum at the resolution considered here, so we fixed it to 4.5. We adopted a single transit depth ($D$), which we would measure for an unspotted star (i.e., a featureless spectrum), although this also can work as a normalization constant for a model spectrum (as we do in Section~\ref{sec:hazes}). We blocked light only from the region described by \tsurf, which we then combined with the spotted region to compute an observed transit depth after normalizing to the combined spectrum with no light blocked (the out-of-transit spectrum). This yielded an observed transit depth at any wavelength, which we convolved with the relevant filter profile (\textit{K2}, MEarth, \textit{HST}, and \spitzer) to compare to the observed data. 
We eliminated one free parameter by fixing \tsurf\ to the temperature assigned from the optical spectrum (3540\,K; Section~\ref{sec:stellar_params}). This assumption was imperfect, as it is possible to reproduce the observed spectrum with significant spot coverage and/or cooler spots while increasing \tsurf.  However, allowing \tsurf\ to go to higher values while forcing it to reproduce the observed spectrum did not significantly change the answer, as we explore further in Section~\ref{sec:spots_spectrum}. 

We compared our model to the observed transit depths data within an MCMC framework using \texttt{emcee} \citep{Foreman-Mackey2013}. Each of the three free parameters evolved under uniform priors, with physical limits ($0<f_S<1$ and $0<D<1$), and those imposed by our model grid (1500\,K$<$\tspot$<$4000\,K). We ran the chain with 12 walkers for 100,000 steps, following a burn-in of 10,000 steps. This was sufficient for convergence based on the autocorrelation time. 

\begin{figure*}[ht]
    \centering
    \includegraphics[width=0.9\textwidth]{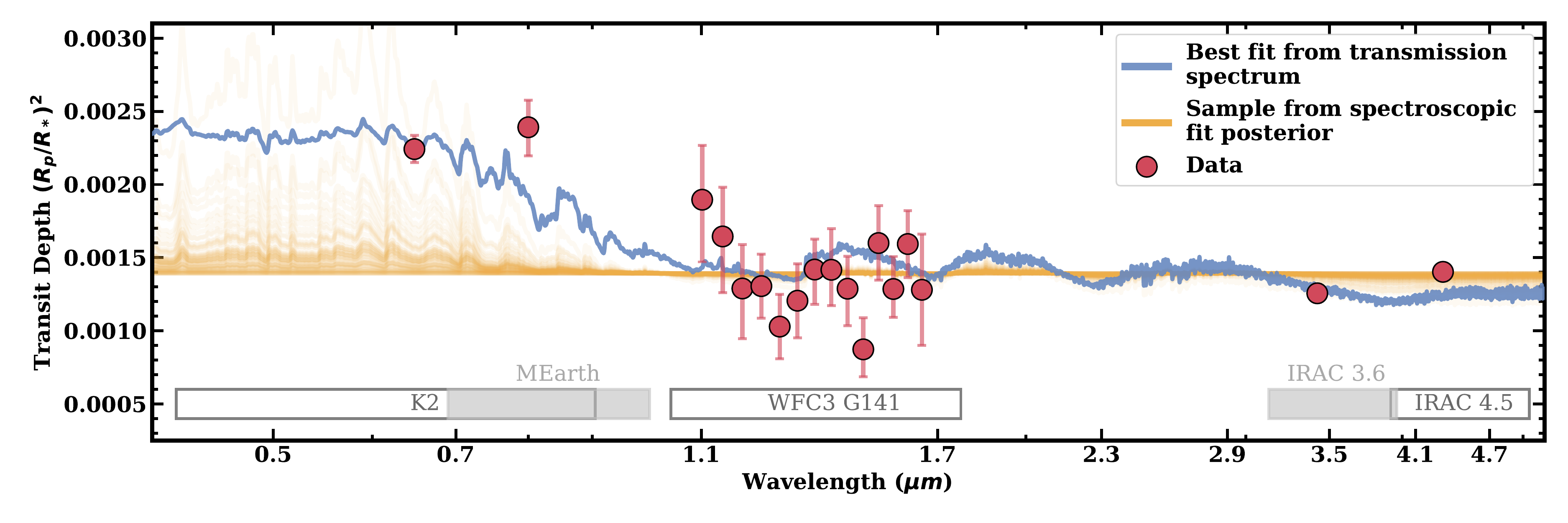}
    \caption{The best-fit model of a planet with a flat transmission spectrum crossing a heavily spotted star along a pristine transit chord (blue). The observed transit depths are plotted in red circles and the labeled bars at the bottom indicate which instruments took those observations. The yellow lines show 100 random samples in \tspot\, $f_S$, and \tsurf\ taken from our fit to the IGRINS spectra (see Section~\ref{sec:spots_spectrum}). None of the results drawn from the spectroscopic constraints are consistent with the data; the only samples that come closest to the optical data fail to reproduce the \spitzer\ data. Results are similar using the posterior from our fit to the SNIFS+SpeX data.
    \label{fig:spots}}
\end{figure*}

\begin{figure}[ht]
    \centering
    \includegraphics[width=0.45\textwidth]{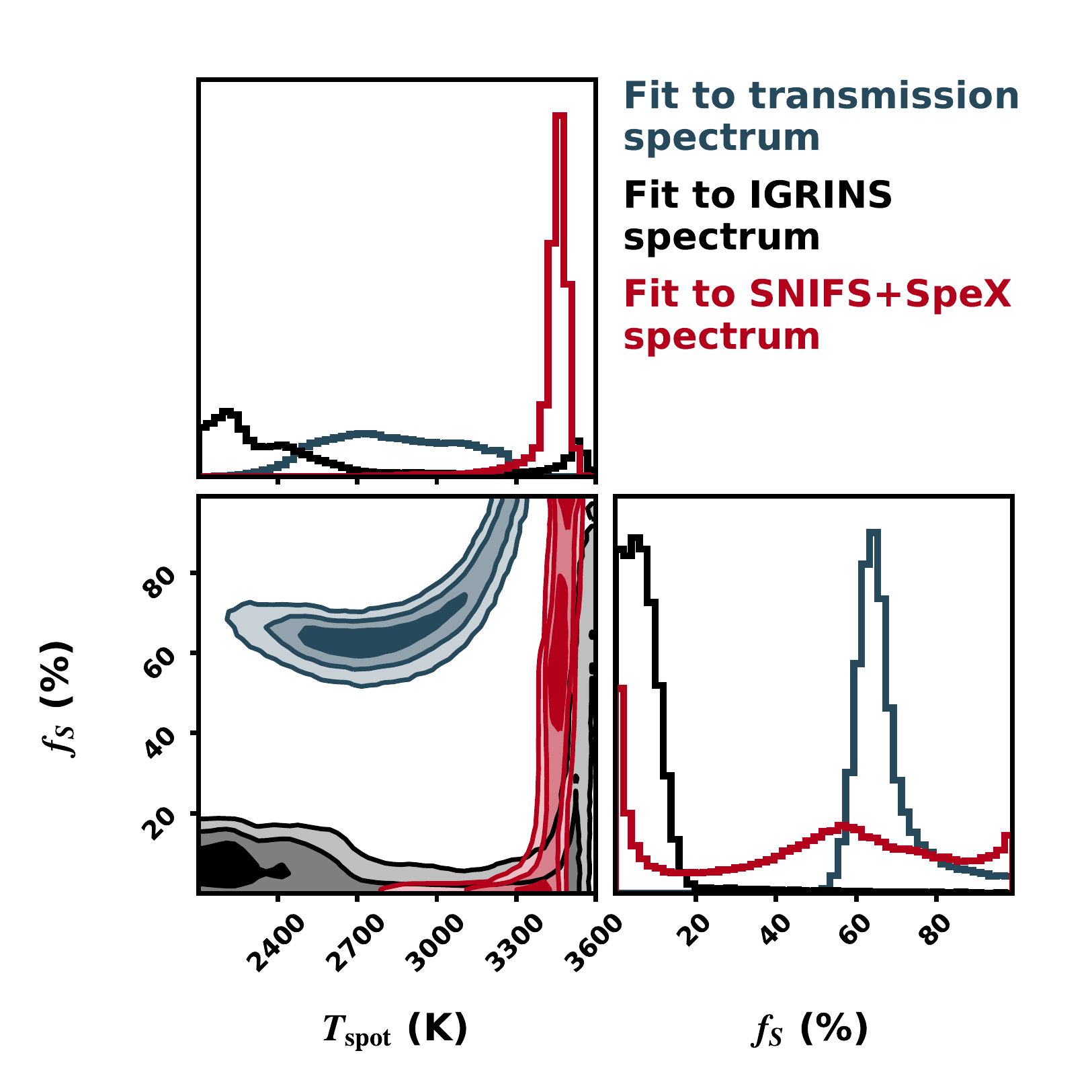} \caption{Corner plot of \tspot\ and $f_S$ from fitting a flat planetary transmission spectrum to the observed transit depths and assuming a pristine transit chord (blue) compared to the same parameters constrained by fitting the stellar spectrum from IGRINS (black) or SNIFS+SpeX (red) spectra with a two-temperature model. Some solutions with low $f_S$ and low \tspot\ values were trimmed for clarity.\label{fig:spots_posterior}}
\end{figure}  

We show the best-fit transmission spectrum in Figure~\ref{fig:spots}, and a subset of the posteriors in Figure~\ref{fig:spots_posterior}. The results suggest we can explain the deeper optical transits if $71\%^{+14}_{-6}$ of the star is covered in spots with \tspot=$2750^{+200}_{-250}$\,K. Such large spot coverage fractions are rare, even for young stars \citep{Morris2020}, but they have been observed on stars in star-forming regions \citep[e.g.,][]{Gully-Santiago2017} so we could not dismiss this option on statistical grounds alone. We explore constraints provided by the transit light curve and stellar spectrum in the next two sections.  

The star is likely to harbor spots with a range of temperatures. However, adding an additional spot with a different temperature and coverage fraction did not alter our conclusions: the preferred solution is 1) to have both spots of similar temperatures, or 2) to have one of the spots be equivalent to the result above, and the other being either too cold or possessing too small a coverage fraction to impact the final result. In either case, the combined effect is not significantly different from the single spot temperature model. The conclusion is still that the spot coverage fraction must be $>60$\% to explain the observed transit depths.

\subsubsection{Spots from the light curve}\label{sec:lightcurve} 

Surface inhomogeneities imprint on the stellar light curve as the star rotates, with spots moving into darker regions of the stellar disk before disappearing from view. All else being equal, larger/more spots and plages will lead to higher-amplitude variation. As discussed in \citet{2021arXiv210200007L}, it is difficult to use this relation in reverse to infer spot coverage fractions from light curves because many different spot coverage fractions and distributions can produce similar light curves. Rather than trying to use the light curve to constrain the spot coverage fraction, we instead attempted to explore what kind of surface patterns could reproduce both the light curve and the observed transit depths.

To this end, we used \texttt{Fleck}\footnote{\url{https://github.com/bmorris3/fleck}} \citep{Morris2020}, which produced light curves from simple limb-darkened spotted stars. We compared the model-generated light curves to the light curve within a MCMC framework using \texttt{emcee}. We restricted our analysis to the \ktwo\ curve; the \hst\ and \spitzer\ curves contain too little out-of-transit data, and the \mearth\ out-of-transit monitoring is not precise enough to add significant constraints above the \ktwo\ data. The \ktwo\ data likely has long-term systematics, which we remove by fitting the light curve with a low-order polynomial before comparing it to the \texttt{Fleck}-generated model.

We fixed the rotation period to that from \cite{Mann2016b} and restricted the spot contrast to be consistent with the results from Section~\ref{sec:lightsource}. Specifically, we convolved the \ktwo\ filter profile with BT-SETTL models (described in Section~\ref{sec:lightsource}) using the assumed \tsurf\ and posterior \tspot\ values. This yielded a spot contrast for every value of $f_S$ (also from the posterior), which we included in our likelihood. The result is that the contrast was fixed based on the model $f_S$ value derived from the spot radii. This also effectively limited $f_S$, as values $\lesssim0.6$ had no corresponding contrast that could reproduce the transmission spectrum. The goal was to determine if the high spot coverage fraction implied by the transmission spectrum could reproduce the light curve (and what would the resulting star look like). 

There were three fit parameters for each spot: latitude, longitude, and radius. Each parameter evolved under uniform priors with physical limits (e.g., latitude from -90 to 90 degrees). We tested using three to seven spots; the results did not change in any way relevant to our analysis. 

\begin{figure*}[hb]
    \centering
    \includegraphics[width=0.95\textwidth]{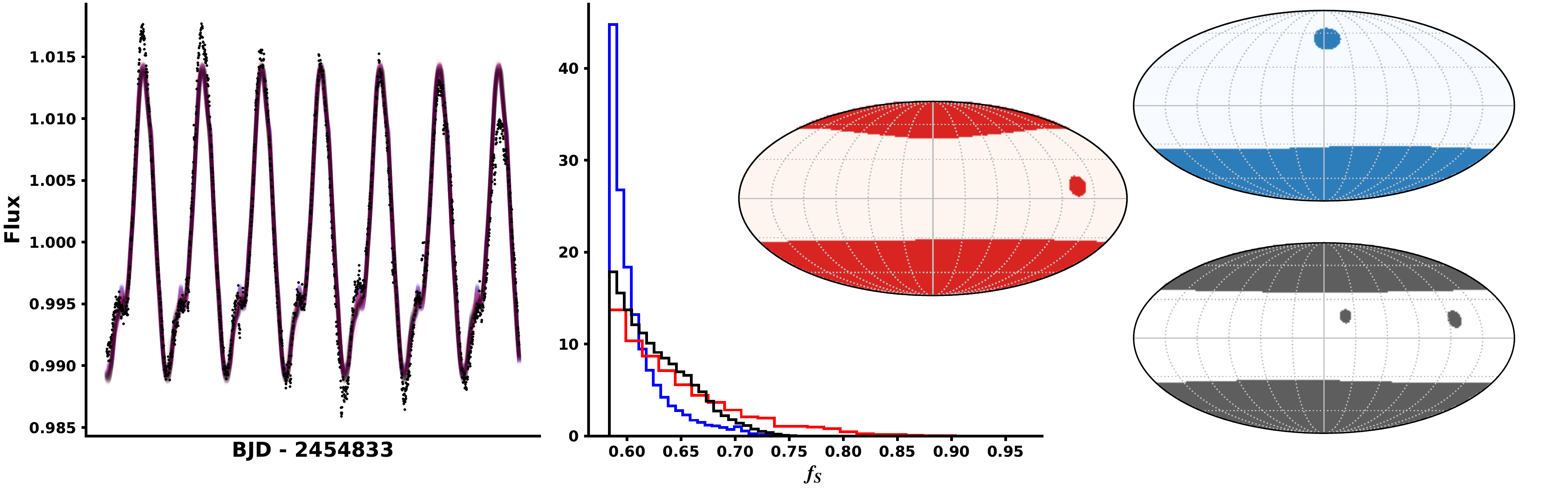}
    \caption{\textit{Left: } Example results from \texttt{Fleck}, showing a portion of the light curve (black points) and best-fit models (purple curves). The model only captures the mean variability, failing to capture the peaks and troughs because it does not include long-term spot evolution. \textit{Middle:} Posteriors of $f_S$. \textit{Right:} Three example stellar surfaces randomly pulled from our fit posteriors. The colors correspond to 2 (blue), 3 (red), and 4 (grey) spots. To explain the low variability in the light curve and the inferred spot coverage, the spots must cover a large portion of the star and be located at the poles. 
    \label{fig:spot_lc}}
\end{figure*}

We highlight some representative results in Figure~\ref{fig:spot_lc}. In order to simultaneously reproduce the low variability ($<2\%$) in the light curve and the inferred large spot coverage, highly symmetric polar spots that cover a large portion of the star are required. Previous studies have suggested the presence of polar spots in M-dwarfs \citep{doyle2018investigating, roettenbacher2018connection} so we can not rule out this possibility. Most often the model prefers either a slight asymmetry in the polar spots and/or 1-3 smaller equatorial spots to explain the perturbations offset from the major sinusoidal pattern seen in the light curve (likely spot clusters at different longitudes). This is consistent with earlier findings that large spot coverage fractions can still produce relatively low variability \cite{johnson2021forward}.

As expected, it was possible to recreate a star that reproduced the light curve and could explain the transit depths using a featureless planetary spectrum; a single-band light curve was not uniquely constraining. However, the resulting stellar surface had to be finely tuned to contain mostly polar spots. In Section~\ref{sec:lightcurve}, we assumed that $f_S$ was not a time variable; while this assumption was not totally correct, the light curve results suggested that it is reasonable nonetheless. The only way to reproduce the low photometric variability was with mostly polar and/or symmetric spots (Figure~\ref{fig:spot_lc}), which, by definition, produce small temporal changes in $f_S$. Or, stated another way, if $f_S$ were changing at $\gtrsim5\%$, this would be evident in the light curve, validating our simplification in Section~\ref{sec:lightcurve}.

\subsubsection{Spot properties from the observed stellar spectrum}\label{sec:spots_spectrum}

\begin{figure*}[hb]
    \centering
    \includegraphics[width=1.0\textwidth]{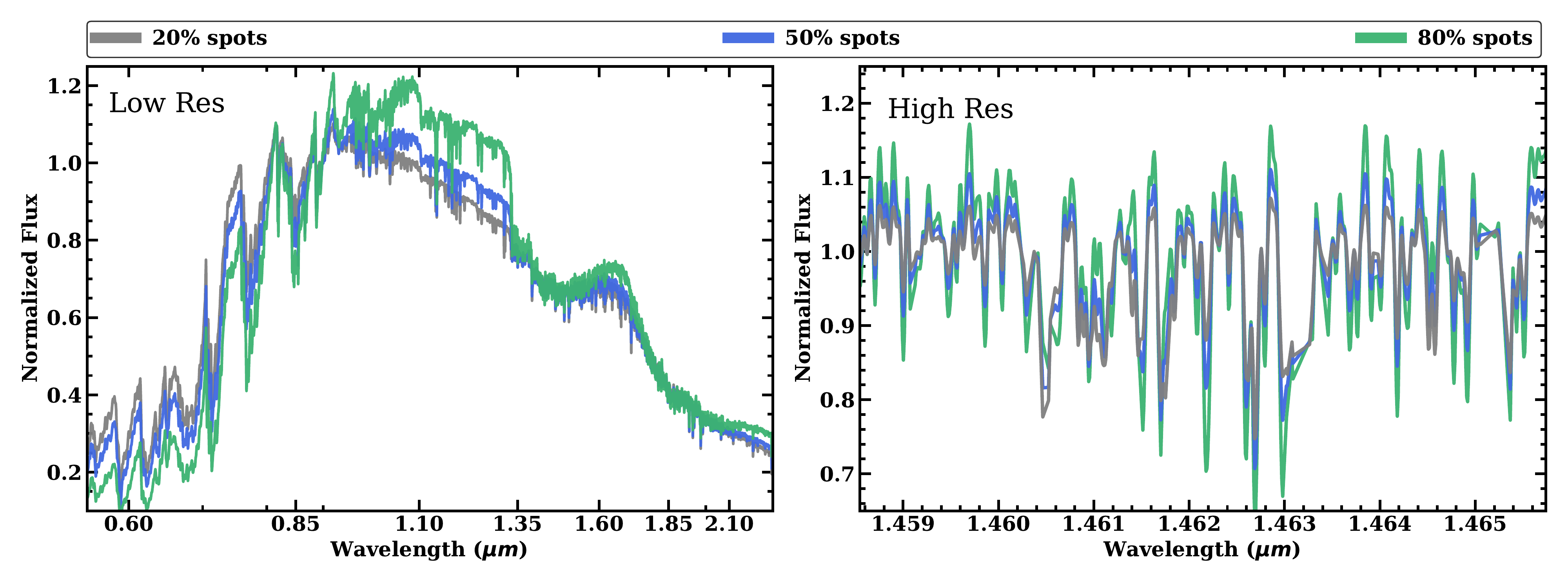}
    \caption{Illustration of the effect of spots on the observed stellar spectrum using BT-SETTL atmosphere models for a star with \tsurf$=3500$\,K and \tspot$=2800$\,K and spot coverage fractions of 80\% (green), 50\% (blue), and 20\% (green). The left panel shows a moderate-resolution optical--NIR spectrum (matching the wavelength coverage and resolution as our observational data). Since spectra only generally provide relative flux measurements, each model is normalized to yield the same total luminosity. The change over the three models is small in the optical, and could largely be fit out by assuming a cooler surface temperature or higher reddening. However, as large spots have a much larger impact on the NIR spectrum, the combined dataset can provide strong constraints on both \tsurf\ and \tspot.
    \label{fig:spot_spectrum}}
\end{figure*}

Large cool spots change both the spectral shape and the strength of molecular features, as we show in Figure~\ref{fig:spot_spectrum}. In extreme cases, cool spots can generate molecular features that should otherwise not be present for a homogeneous surface. Numerous earlier studies have taken advantage of this to study spot properties of young stars \citep[e.g.,][]{1994A&AS..107....9P, 2016MNRAS.463.2494F}, likely providing more accurate pictures of stellar surfaces than light curve based metrics alone \citep[e.g.,][]{Gully-Santiago2017}. 

Since the effect of spots on the observed spectrum is strongly wavelength dependent (e.g., the cooler region has a smaller overall contribution in the optical), it can be detected in spectra with broad wavelength coverage. The optical data mostly probes \tsurf\ and reddening, while the NIR data is more sensitive to changes in \tspot\ and $f_S$. We used the moderate resolution ($R\simeq1000-2000$) spectra from \citet{Mann2016b} to explore the spot properties for K2-33. The spectra were built from two overlapping datasets, one from the SNIFS spectrograph \citep[0.32--0.95\um; ][]{Lantz2004} and one from the SpeX spectrograph \citep[0.7--2.4\um; ][]{Rayner:2003}. Both spectrographs provide excellent relative flux calibration \citep[2-4\%;][]{Rayner2009, Mann2013c} ideal for this work. We refer readers to the discovery paper \citep{Mann2016b} for more details on the observations and reduction of these data.

We fit the spectra with a two-temperature model \citep[e.g., ][]{Gully-Santiago2017, 2019AJ....158..101M} derived from BT-SETTL atmosphere models. In total, we had 13 free parameters for the fit. The first three, \tsurf, \tspot, and $f_S$ were as described in Section~\ref{sec:transit_fitting}, but without any limit on \tsurf\ and an additional requirement that \tspot\ be at least 50\,K less than \tsurf. Six additional parameters described the normalization and slope of the blue-optical, red-optical, and NIR data ($a1$, $a2$, $b1$, $b2$, $c2$, and $c2$). These were needed because the SNIFS blue end, SNIFS red end, and SpeX data were calibrated separately (SNIFS arms are separated with a dichroic at 0.54\um); as a result, the three regions are often offset at the 2-5\% level even after calibration \citep{Mann2015a}. Two additional parameters handle small wavelength offsets between the data and models, one for the optical/SNIFS data ($\lambda_{\rm{off,opt}}$) and one for the NIR/SpeX data ($\lambda_{\rm{off,NIR}}$). The remaining two parameters were extinction ($A_V$) and a parameter to handle underestimated uncertainties or other systematics in the spectra ($\sigma_f$), which we treated as a fractional error added in quadrature with measurement errors. 

Following \citet{Mann2013c}, we masked specific regions where the models poorly reproduced observed spectra of stars with empirically-determined temperatures (from long-baseline interferometry). No other direct constraints were put on the data.

All parameters were fit using \texttt{emcee}. We placed Gaussian priors on the slope parameters ($a2$, $b2$, $c2$) based on estimates of the flux calibration from \citet{Mann2013c} and \citet{Rayner2009}. All other parameters evolved under uniform priors. We ran the MCMC chain with 100 walkers for 500,000 steps, which was more than sufficient for convergence based on the autocorrelation time. 

We added the resulting $f_S$ and \tspot\ distributions in Figure~\ref{fig:spots_posterior}. As expected, spot temperatures were unconstrained when the spot fraction was low, and the spot fraction was unconstrained when \tspot\ was close to \tsurf. The general result was that either spots must be relatively warm $\gtrsim3000$\,K or cover a small fraction ($<10$\%) of the star. 

We performed a similar fit using the high-resolution IGRINS spectra from \citet{Mann2016b}. IGRINS covers most of the $H$- and $K$-bands with a resolution of $\simeq$45,000. Nearly-identical data were used effectively in \citet{Gully-Santiago2017} to study the spot properties of a young star. Of the 54 IGRINS orders, we focused on the four with the highest SNR, lowest telluric contamination, and highest sensitivity to spots (e.g., right panel of Figure~\ref{fig:spot_spectrum}). The fit parameters were similar to what we used for the moderate-resolution data, but each order had separate flux calibration terms, a common wavelength offset, and we did not fit for $A_V$ (which has a minor impact on the NIR data). We add an additional parameter ($b$), which accounts for both rotational and instrumental broadening as a single Gaussian width. All parameters were fit under uniform priors, with \texttt{emcee} using 100 walkers and 100,000 steps. 

The results from the IGRINS orders were broadly consistent with that from our SNIFS+SpeX data; the fit strongly prefers either \tsurf$\simeq$\tspot\ or $f_S<20\%$. The major differences from our SNIFS+SpeX results were that the IGRINS fit preferred a warmer \tsurf\ (and hence warmer \tspot) and that the fit to IGRINS data allowed for lower temperature ($<2500$\,K) spots with higher surface coverage fractions (up to $\simeq$20\%) than the fit to the SNIFS+SpeX data. 

A comparison between the posteriors from the fit to the stellar spectrum and the fit to the transmission spectrum (Figure~\ref{fig:spots_posterior}) had no overlap. The fit to the stellar spectrum rules out all possible \tspot\ and $f_S$ combinations that could explain the observed transmission spectrum. This was true regardless of the dataset used. 

While the SNIFS+SpeX results are more precise, we used the fit from IGRINS for further analysis. The main reason was that tests on similar spectra (same instrument and setup) for other stars suggested that the SNIFS+SpeX spectra were sensitive to the relative flux calibration. While both SNIFS and SpeX have excellent flux calibration \citep{Mann2015b}, K2-33 was observed at an air mass of $>1.4$, around which the blue end of the (SNIFS) data become less reliable \citep{Buton2013}. Further, IGRINS data was taken over many epochs (yielding similar results regardless of which IGRINS spectrum we used) and the data was contemporaneous with the MEarth and \spitzer\ transits. Overall, we consider the IGRINS fit to be both more accurate and more conservative.

We show a more fair comparison of the spectroscopic results to the transit data in Figure~\ref{fig:spots}. For this, we randomly drew \tsurf\ and \tspot\ values from the (stellar spectrum fit) posterior and computed the predicted transit depth as a function of wavelength, normalized to the depth at 4.5\um. As can be seen, even the most extreme samplings yielded transit depths inconsistent with the \mearth\ and \ktwo\ data. 

For our fit to the transmission spectrum, we had one fewer free parameter, as we fixed \tsurf\ to 3540\,K; for our fit to the stellar spectrum, both \tsurf\ and \tspot\ were free parameters. Fixing \tsurf\ in the spectroscopic fit or shifting the posterior on \tsurf\ (for the IGRINS fit) did not change the conclusion: the spectroscopic fit was still inconsistent with the spot coverage required to explain the observed transit depths.

The uncertainties from both fits were likely underestimated due to imperfect models, but this is also unlikely to impact our results. For example, we tried using the PHOENIX model atmospheres from \citet{2013A&A...553A...6H} which yielded \tsurf\ and \tspot\ values $\simeq60$\,K cooler and $A_V$ values 0.1 magnitudes lower. These changes were larger than the errors within a fit, and are likely due to differences in the handling of strong molecular lines in the spectra of cool stars \citep{Mann2013c, 2016A&A...587A..19P}. However, the conclusions on $f_S$ were largely unchanged; \tspot\ must either be within $\simeq150$\,K of \tsurf\ or $f_S$ must be below $\simeq$20\%. This was expected: these model systematics impact both the surface and spot spectra in similar ways, keeping the difference in temperature similar. We concluded that even in the presence of significant systematics in the model of M dwarfs, the spectrum rules out any spot pattern that can explain the transit depths. 

It is likely that the stellar surface contains more than a single spot, and hence more than a single spot temperature. However, adding additional spots with different \teff\ values did not meaningfully change the conclusion. The additional spot simply followed the output distribution in \tspot\ and $f_S$ of the single spot assumed above, but with larger uncertainties. It is possible that allowing an arbitrary number of spots each with unconstrained temperatures can reproduce the observed transmission spectrum. However, our tests suggest that such a reconciliation would come primarily by increasing the uncertainties until the contours had marginal overlap; in this case, the fit does not improve meaningfully with more spots. 

\section{Photochemical Hazes} \label{sec:hazes}

In addition to spots, the existence of submicron aerosol particles in K2-33b's atmosphere could also lead to enhanced optical transit depths. One possible formation pathway for such particles is photochemistry resulting in the breakup of small molecules at low pressures ($\sim$1 $\mu$bar) and subsequent polymerization of photochemical products, forming high altitude hazes \citep[e.g.][]{morley2015, kawashima2019b, gao2021rev}. The small size of these particles causes a rapid decrease in opacity towards longer wavelengths and thus a decrease in transit depth. This effect is strengthened in young planets like K2-33b that may be experiencing atmospheric outflows, which are capable of pushing haze particles to greater altitudes and increasing the difference in transit depth between the optical and the NIR \citep{Wang2019, Gao2020, ohno2021}. 

We explore the impact of hazes on K2-33b's transmission spectra following the method described in \citet{Gao2020}. Briefly, we construct model atmospheres defined by user-selected core masses and atmospheric mass fractions composed of a convective zone at depth and a radiative region at lower pressures separated by the radiative-convection boundary, where the temperature is set to the equilibrium temperature of the planet assuming zero albedo and full heat redistribution; we use the Rosseland mean opacity from \citet{freedman2014} to find the radiative-convective boundary and to calculate the temperature-pressure profile in the radiative zone. The core is assumed to have Earth-like composition, with the mass-radius relationship from \citet{zeng2019}. As an update to \citet{Gao2020}, we use a planetary evolution model \citep{tang_aas,lopez2014} to derive an intrinsic temperature as a function of core mass, yielding values $\sim$100-150 K for core masses of 3-10$M_{\oplus}$, atmospheric mass fractions $<$10\%, and model planet ages matching that of K2-33b. 

We consider the effect of atmospheric loss pushing haze particles upwards. We compute atmospheric loss rates assuming energy limited escape with an energy-deposition pressure level of 1 nbar and a mass-loss efficiency of 10\% \citep[e.g.][]{lopez2017}. We use the saturated stellar XUV flux for M dwarfs \citep{wright2011, wright2018, pineda2021} to drive the atmospheric loss. 

The elemental composition of the atmosphere is simplified to just H, He, O, and C, with solar abundances of O and C incorporated into water vapor and methane, respectively, and the rest of the atmosphere composed of H/He. We initially choose methane as the dominant carbon carrier (``\ce{CH4} models'') due to the relatively low temperature of K2-33b \citep{lodders2002}, but we also explore the impact of upward mixed \ce{CO} (``\ce{CO} models'') on the transmission spectra \citep[see][and references therein]{fortney2020}. Given this atmospheric composition and assumed fully mixed abundance profiles, we equate the haze production rate profile to the methane photolysis rate profile multiplied by a haze production efficiency factor, which is a free parameter. The methane photolysis rate profile is generated using a simplified photochemical scheme where methane is photolyzed by Lyman $\alpha$ radiation from the host star, with an assumed Lyman $\alpha$ flux of a saturated M dwarf \citep{pineda2021}. 

Once produced, haze particles are transported to the deep atmosphere via sedimentation and mixing, the latter of which is parameterized using eddy diffusion. The eddy diffusion coefficient is a free parameter and assumed to be constant with altitude. The haze particles can grow through coagulation during transport, starting from a fixed minimum radius of 10 nm, and we assume that all haze particles are spherical. The haze evolution is simulated using the Community Aerosol and Radiation Model for Atmospheres \citep[CARMA;][]{turco1979,toon1989,ackerman1995,gao2018}, which has been applied to exoplanet and solar system hazes on numerous occasions \citep{gao2017,adams2019,Gao2020,gao2020hj}. We use Mie scattering to calculate the haze optical properties and consider two haze compositions: soot and tholin, with the corresponding complex refractive indices \citep{lee1981,khare1984,chang1990,morley2015,gavilan2016,lavvas2017}. Finally, we combine the model atmosphere and haze optical properties to compute the model transmission spectra.

\subsection{\ce{CH4} models}

Since the mass of the planet is unknown, we considered several planet core masses. From the planet mass-radius distribution, we calculated a predicted mass for K2-33b of $16.59_{-11.73}^{+7.08} M_{\oplus}$ using the python package, \texttt{forecastor} \citep{ChenKipping2017}. Due to the youth of the planet, we expect the true value to be smaller because younger planets are larger and less dense than their evolved counterparts \citep{owen2020constraining}. As such, we tested core masses of 3, 5, and 8$M_{\oplus}$. We also examined eddy diffusion coefficients of 10$^{9}$ and 10$^{11}$ cm$^2$ s$^{-1}$ and haze production efficiencies of 1, 2, 4, and 8\% \citep{lavvas2017}. For each core mass, we varied the atmospheric mass fraction to roughly match the observed planet radius, resulting in values of 1--2.5\%. 

To compare our models to the data, we convolved the model spectra with the relevant filter profiles to create synthetic transit depths corresponding to each effective wavelength (photon weighted mean wavelength). The effective wavelength factors in the widths of the broadband filters were calculated using K2-33b's spectrum and each filter's bandpass. The results of this calculation yielded effective wavelengths of 0.72 $\mu$m (\textit{K2}), 0.84 $\mu$m (MEarth), 3.46 $\mu$m (Channel 1),  4.43 $\mu$m (Channel 2). We added a free parameter, a normalization constant, to allow each model spectra to shift in median depth, and varied it to minimize the $\chi^{2}$ when compared to our data. Although the \citet{Gao2020} model computes the radius of the model planets, we allow this normalization factor to shift the planet radius to account for model uncertainties in e.g., the intrinsic temperature, atmospheric composition, etc. 

Our results disfavored a featureless model and all haze models where the carbon carrier in the background atmosphere is \ce{CH4} ($>4 \sigma$ confidence) regardless of the haze composition, core mass, atmospheric mass fraction, eddy diffusion coefficient, and haze production efficiency. The lowest $\chi^{2}$ for these models that had a normalization factor within 20\% was $\chi^{2} =$ 58 for tholin hazes and $\chi^{2} =$ 79 for soot hazes. These high values are driven by the disagreement between the models and the \textit{Spitzer} points. The normalization parameters and the corresponding $\chi^{2}$ values are listed in Table \ref{tab:models_chi_square}. Only the top three lowest $\chi^{2}$ values are listed for each haze composition.

\subsection{\ce{CO} models}

Under the assumption of thermochemical equilibrium and solar metallicity, methane should be the dominant carbon carrier in the atmospheres of planets with K2-33b's equilibrium temperature \citep{lodders2002}. However, no methane has been detected in the atmospheres of exoplanets that are similar in temperature, but older  \citep{kreidberg2018,benneke2019}. One possible reason for this is the disequilibrium process of mixing bringing \ce{CO}-rich gas from depth to the observable part of the atmosphere \citep{fortney2020}. To test this scenario and better fit the \textit{Spitzer} points, we replace \ce{CH4} in our previous models with \ce{CO} and reduce the water vapor abundance accordingly. The same core masses, atmospheric mass fractions, eddy diffusion coefficients, and haze production efficiencies were considered in generating the \ce{CO} model grid. We use the same haze production rate even though it is computed assuming methane photolysis so as to isolate the impact of changing atmospheric composition on the transmission spectrum. Given the unknowns in the haze production pathway and experimental evidence pointing to \ce{CO} as a possible haze precursor \citep{horst2018,he2018,fleury2019}, this simplification is justified.

Table \ref{tab:models_chi_square} lists the top 3 lowest $\chi^{2}$ and normalization factors for each haze composition. Our results disfavor hazes composed of soot to $>5\sigma$ confidence, as they are unable to fit the optical data. In contrast, tholin hazes ($\chi^{2} = 29$, \textit{dof}=16) give reasonable fits to our measured transmission spectrum ($>2 \sigma$ confidence).

To summarize, consideration of photochemical hazes allowed us to put the following constraints on our model parameters: 
\begin{itemize}
\item \textit{planet core mass}: Lower values (3 or 5$M_{\oplus}$) preferred.  
\item \textit{atmospheric mass fraction:} 1--2.5\% preferred.
\item  \textit{eddy diffusion coefficient}: Higher values (10$^{11}$ cm$^2$ s$^{-1}$) preferred for tholin hazes, while lower values (10$^{9}$ cm$^2$ s$^{-1}$) preferred for soot hazes.
\item \textit{haze efficiencies}: Soot models prefer lower values (1\%), while tholin models prefer higher values (4\%)
\item \textit{haze composition}: Tholin strongly preferred over soot.
\item \textit{primary carbon carrier}: \ce{CO} strongly preferred over \ce{CH4}.
\end{itemize}

The best-fit model from our grid has a 5$M_{\oplus}$ core and an atmospheric mass fraction of 2.2--2.3\%. The model atmosphere possesses a high (10$^{11}$ cm$^2$ s$^{-1}$) eddy diffusion coefficient, a haze production efficiency of 4\% \citep[about midway between that of Jupiter and Titan;][]{lavvas2017}, and tholin hazes, with \ce{CO} as the dominant gaseous carbon carrier. 

\begin{deluxetable*}{lcccccr}
\tablecaption{Normalization Factor for Atmospheric Models  \label{tab:models_chi_square}}
\tablehead{
\colhead{Haze} &
\colhead{Planet Core Mass} &
\colhead{Atmospheric Mass } &
\colhead{Eddy Diffusion} &
\colhead{Haze Production} & 
\colhead{Normalization} & 
\colhead{$\chi^2$}\tablenotemark{a} \\
\colhead{Composition} & 
\colhead{[$M_{\oplus}$]} &
\colhead{Fraction} & 
\colhead{Coefficient [cm$^2$ s$^{-1}$]} & 
\colhead{Efficiency [\%]} &
\colhead{Factor} & 
\colhead{} } 
\startdata
\multicolumn{7}{c}{\emph{Haze Fit with No Spots}}\\
\hline
&   5.0 & 0.024 & 10$^{11}$ &    4.0 &      0.87 &         58 \\
\ce{CH4} Tholin  &   5.0 & 0.025 & 10$^{11}$ &    4.0 &      0.81 &         58 \\
&   5.0 & 0.023 & 10$^{11}$ &    4.0 &      0.93 &         59 \\
\hline
&   3.0 & 0.013 & 10$^{9}$  &    1.0 &      0.84 &         79 \\
\ce{CH4} Soot &   3.0 & 0.012 & 10$^{9}$ &    1.0 &      0.80 &         80 \\
&   3.0 & 0.013 &10$^{11}$  &    1.0 &      0.86 &         84 \\
\hline
\hline
&   5.0 & 0.022 & 10$^{11}$ &    4.0 &      1.07 &         29 \\
\ce{CO} Tholin &   5.0 & 0.023 & 10$^{11}$ &    4.0 &      1.00 &         29 \\
&   5.0 & 0.021 & 10$^{11}$ &    4.0 &      1.14 &         30 \\
\hline
&   3.0 & 0.013 & 10$^{9}$ &    1.0 &      0.88 &         65 \\
\ce{CO} Soot  &   3.0 & 0.014 & 10$^{9}$ &    1.0 &      0.83 &         68 \\
&   3.0 & 0.015 & 10$^{9}$ &    1.0 &      0.80 &         68 \\
\hline
\hline
Flat Line & $\dots$  & $\dots$ & $\dots$ & $\dots$ & $\dots$ & 122 \\
\hline 
\hline 
\multicolumn{7}{c}{\emph{Haze Fit with Spots}}\\
\hline
& 5.0 & 0.020 & 10$^{11}$ & 4.0 & 1.01 & 28 \\ 
\ce{CO} Tholin & 5.0 & 0.021 & 10$^{11}$ & 4.0 & 0.98 & 28\\ 
& 5.0 &	0.019 & 10$^{11}$ & 4.0 & 1.04 &	28\\ 
\hline
& 3.0 & 0.010 & 10$^{11}$ & 1.0 & 1.12 & 56\\ 
\ce{CO} Soot & 3.0 & 0.011 & 10$^{11}$ & 1.0 &  1.00 & 59 \\ 
& 3.0 &	0.012 & 10$^{11}$ &	2.0	& 0.80 & 61 \\ 
\enddata
\tablenotetext{a}{The spot-free haze fits has a \textit{dof}= 16 and the combined haze and spot fits has a \textit{dof}=14.}
\tablecomments{Only the top 3 (lowest $\chi^{2}$ values) for each haze composition is listed here. In addition to this, for the combined haze and spots fits, the fit also needed to yield spot properties consistent with the fit to the stellar spectrum. }
\end{deluxetable*}

\begin{figure*}[ht]
    \centering
    \includegraphics[width=0.9\textwidth]{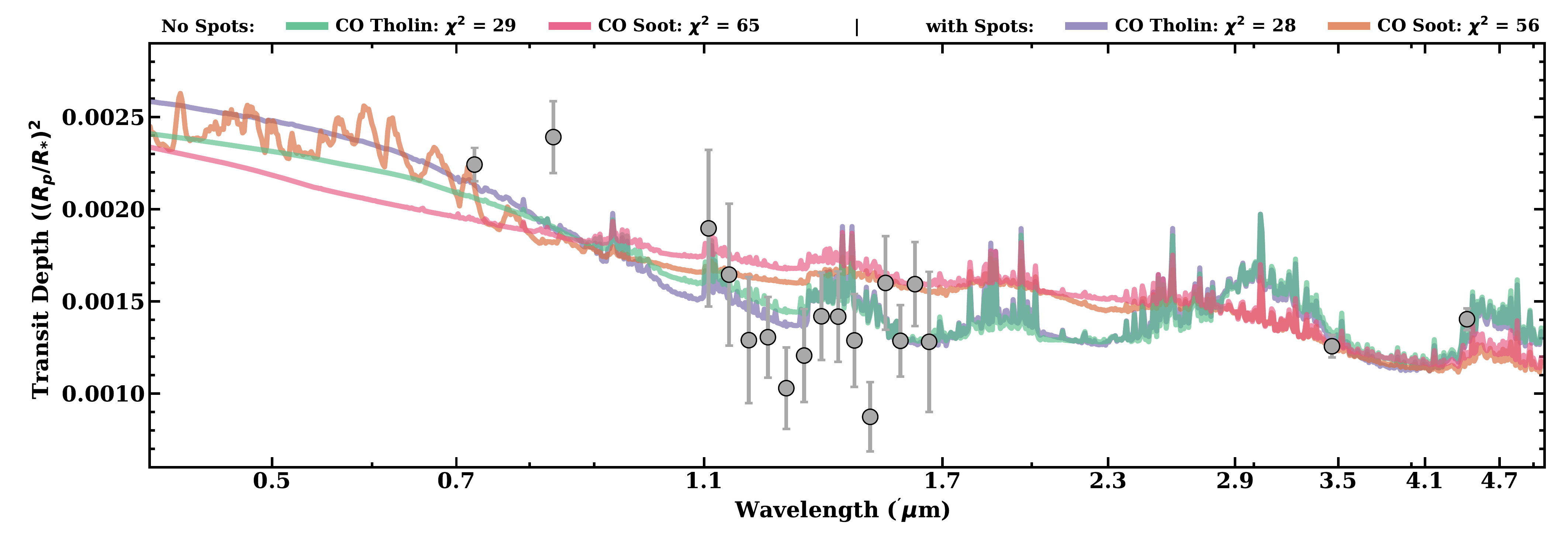}
    \caption{Transmission spectrum of K2-33b from our data (gray circles) compared to haze models composed of 1) \ce{CO} Tholin (green), 2) \ce{CO} Soot (pink), 3) a combined spot model and \ce{CO} Tholin (purple), and 4) a combined spot model and \ce{CO} Soot (orange). Only the models that produced the lowest chi-square values are shown. All models were normalized to give the best fit to the data. The combined haze and spot models gave a lower chi-square value. All calculations were done with high-resolution models. \label{fig:hazes}}
\end{figure*}

\begin{figure}[ht]
    \centering
    \includegraphics[width=0.5\textwidth]{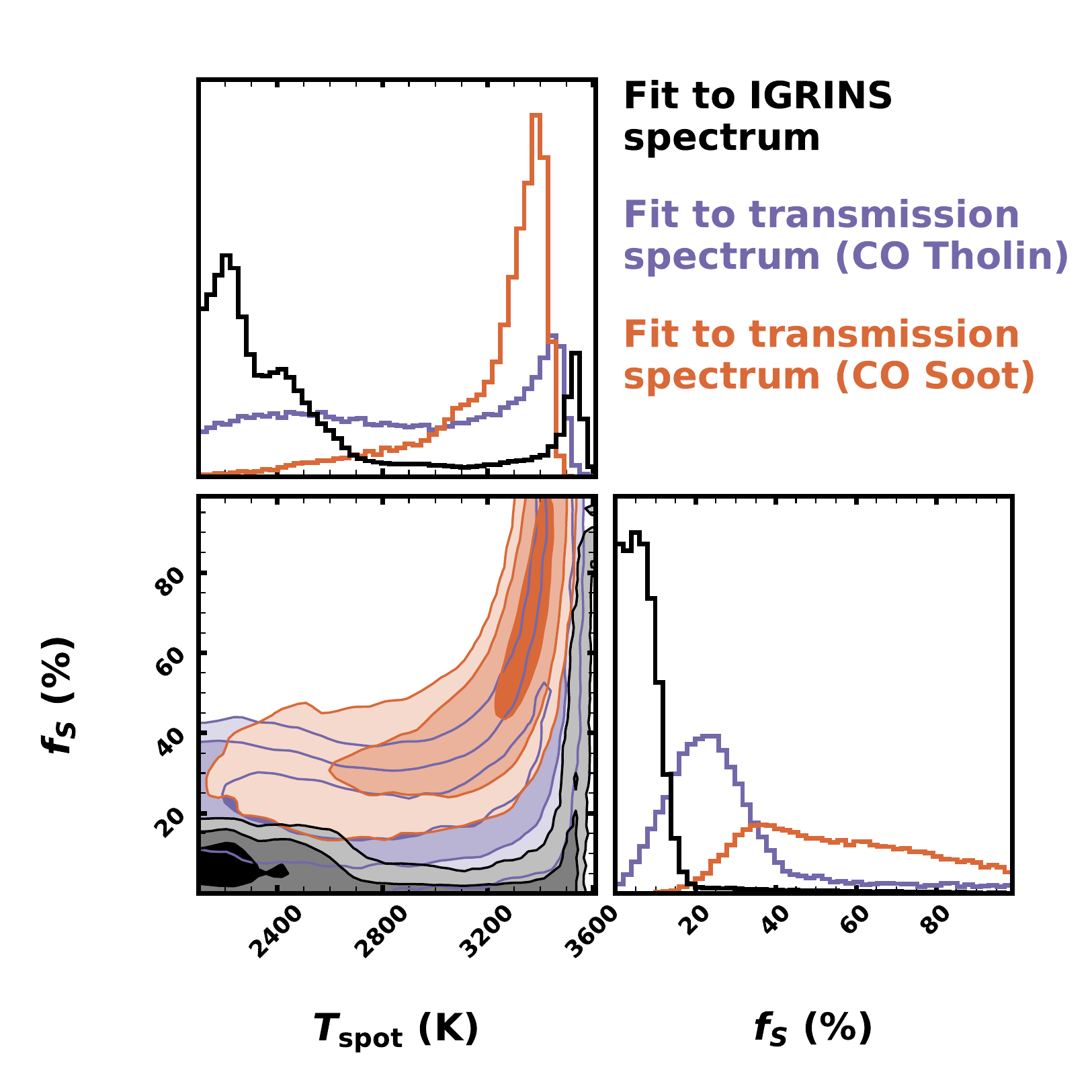}
    \caption{ Posteriors of spot properties from the stellar spectrum (black/grey), a fit to the transmission spectrum using \ce{CO} Tholin models (purple), and a fit to the transmission spectrum using \ce{CO} soot models (orange). The two transmission spectrum fits shown here are those using the Tholin or Soot model with the lowest $\chi^2$ that also yielded spot properties consistent with the fit to the stellar spectrum.  \label{fig:hazes_spots_corner}}
\end{figure}

\subsection{Combining spots and hazes}\label{sec:spotandhaze}

As we showed in Figure~\ref{fig:spots}, spots alone did a relatively poor job fitting the overall transit depths, and the best fits required spot properties inconsistent with the stellar spectrum. Photochemical hazes provided a better fit to the data, but there is still some tension. Here, we endeavored to improve the fit by combining hazes and spots. 

Our methodology was the same as described in Section~\ref{sec:lightsource}, except we replaced the flat model with the theoretical haze models from Section~\ref{sec:hazes} and $D$ was changed to a dimensionless scale factor applied to the models (denoted $D_{\rm{mod}}$). The scale factor can be an arbitrary number, but changing the planet size $>$ 20 \% invalidates the model assumption, so we limited the scale factor to $0.8<D_{\rm{mod}}<1.2$; more extreme scaling would imply a different surface gravity and hence invalidate the model atmosphere. The two other parameters (\tspot\ and $f_S$) evolved under uniform priors. For each model, we ran with 50 walkers and as many steps as required to pass 50 times the autocorrelation time (checking every 1000 steps). The number of steps required varied from 4,000 to more than 15,000, depending on the model.

As with the fits ignoring effects from stellar spots, the models with \ce{CO} as the primary carbon carrier and tholin hazes gave the best fit. This is because the \ce{CO} models are able to match the shallower transit at 3.6\um\ compared to that at 4.5\um. Spots alone cannot explain this difference, as their effect is weak past 3\um\ and even strong spots tend to predict a deeper transit at 3.6\um\ instead of a shallower one. 

Overall, adding spots provided only marginal improvement when fitting the transmission spectrum. For example, the best fit haze model (both with and without spots) was the \ce{CO} model with a tholin haze (which we show in Figure~\ref{fig:hazes}), with an improvement of $\Delta\chi^2\simeq1$ over a spot-free haze model. Given that there were two additional free parameters in the spot and haze fit (compared to haze-only), this improvement was not significant for most of the tholin models (see Table~\ref{tab:models_chi_square} for a summary of the best fits). However, it is worth highlighting that the joint fit of the spots and \ce{CO} Tholin model makes everything consistent: the fit to the stellar spectrum, the fit to the light curve, and the fit to the transmission spectrum. The required small spot coverage fractions and spot properties were consistent with the stellar spectrum (Figure~\ref{fig:hazes_spots_corner}).

Including the effect of spots had a larger improvement for fits using the soot models, with the best cases going from $\chi^2\simeq65$ to $\chi^2\simeq$35. However, these generally required spot coverage fractions and temperatures inconsistent with those inferred from the stellar spectrum. If we consider just cases where the required spot properties were consistent with the stellar spectroscopic data, the best fit soot haze and spot combination did a poor job reproducing the full transmission spectrum ($\chi^2=56$).

\section{Summary and Discussion}\label{sec:summary}
To explore the transmission spectrum of the $\simeq$10\,Myr, K2-33b, we combined 33 transit observations taken by \textit{K2}, MEarth, \textit{HST}, and \spitzer\ spanning over $>$ 2.5 years. We also updated the stellar parameters based on the parallax from \gaia. 

\begin{figure}[ht]
    \centering
    \includegraphics[width=0.5\textwidth]{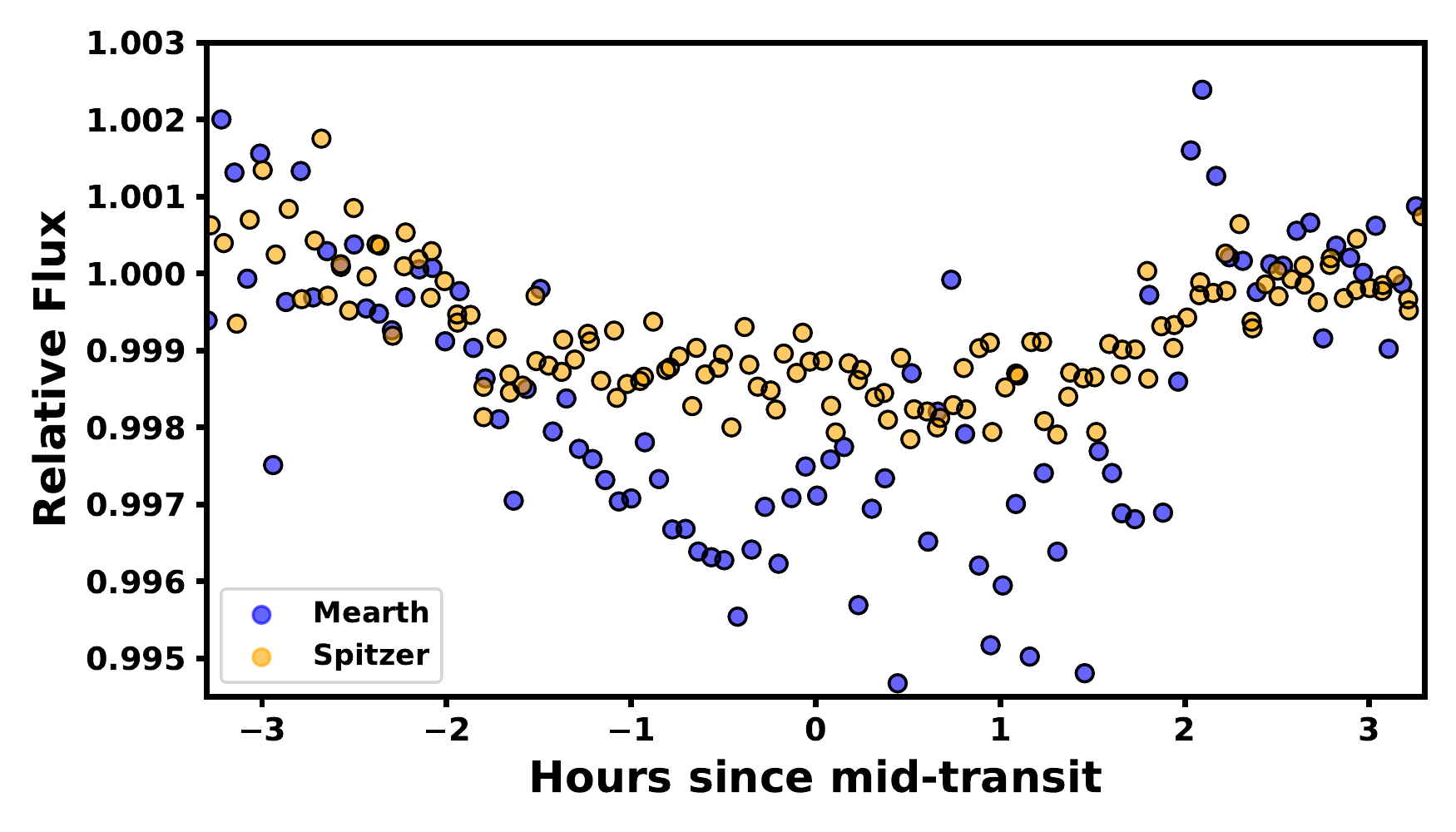}
    \caption{Light curves from \mearth\ and \spitzer{} for the three transits with overlapping data in both (transits 186, 188, and 191). Data from both facilities were  folded and (median) binned for clarity. One of the \mearth\ transits was partial (transit 188) and conditions varied both between and during \mearth\ transits leading to inconsistent data quality. However, the \mearth\ light curve still clearly favors a deeper transit depth. 
    \label{fig:simultaneous}}
\end{figure}

The most striking result from the multi-wavelength data was that the optical transits from \ktwo\ and \mearth\ are almost a factor of two deeper than the NIR transits from \hst\ and \spitzer\ (0.24\% vs. 0.13\%). We explore whether this depth difference is due to unconstrained systematics in the datasets or if it is astrophysical in nature. We rule out the first scenario primarily because the difference holds across multiple datasets, with roughly consistent (but shallower) depths from \spitzer\ and \hst\ and consistent (but deeper) depths from \ktwo\ and MEarth. Agreement between instruments rules out issues related to the data source, data quality, or analysis method. The \spitzer, MEarth, and \ktwo\ data each cover at least five transits and span numerous rotational cycles, ruling out single events like flares. Further, three transits were observed simultaneously by MEarth and \spitzer\ (transit numbers 186, 188, and 191; Table~\ref{tab:obslog}). Even considering just these three transits, the \mearth\ and \spitzer\ depths were still clearly inconsistent (Figure~\ref{fig:simultaneous}). This rules out long-term evolutionary effects. 

In our analysis, we excluded two \spitzer\ and two \mearth{} transits. The two \mearth\ transits were removed primarily because of poor coverage (insufficient out-of-transit baseline). Including them had negligible effect on the transit depth because the GP was too poorly constrained. As we show in Figure~\ref{fig:rprs_spitzer}, the two excluded \spitzer\ transits favored deeper transits, although Channel 1 data this was due to a visible flare before the transit. As a test, we repeated our fit using \texttt{MISTTBORN} including all  transits but masking out the data right around the flare (in AOR 60658432). Otherwise, the analysis was identical to that described in Section~\ref{sec:transit_fit_spitzer}. For channel 1, this gave $R_P/R_*=0.0358\pm0.0010$ and for channel 2, $R_P/R_*=0.0383\pm0.0012$. Neither of the two fits were inconsistent with our original analysis and both were still $>5\sigma$ inconsistent with the depth from \ktwo. 

We conclude that the depth difference is astrophysical. We consider two causes: (dark) spots on the stellar surface, or hazes in the planet's atmosphere -- both of which predict deeper transits in the optical. In the case of star spots, their effect is largest at the optical wavelengths because the contrast ratio of stellar spots decreases with increasing wavelength. As for the hazes, their small particle size results in a decrease in opacity with increasing wavelength, and thus a decrease in transit depth. 
 
We tested three scenarios; 1) a spotted star with a (perfectly) flat planetary transmission spectrum, 2) a mix of photochemical hazes with a pristine star, and 3) a mix of both spots and hazes. Here is a summary of each fit: 
\begin{itemize}
    \item \textit{Spots: } The deeper optical transits can be explained if $>$60 \% of the star is covered in cold ($<$3200 K) spots. However, such large spot coverage fractions are ruled out by the stellar spectrum, which indicate the spots must either cover $<20\%$ of the surface or be similar in temperature to the surface ($>$3,300 K).

    \item \textit{Photochemical Hazes: } A planet with a low core mass ($\simeq5M_\oplus$) and a tholin haze provides the best fit to the data ($\chi^2=29$ with 16 degrees of freedom). All soot models provided a poorer fit than tholin models. This is because tholins provide more absorption in the blue optical and UV than in the NIR, resulting in a steeper optical slope, whereas soots are more effective at absorbing at all wavelengths leading to a more gentle slope (Figure \ref{fig:hazes}). For any haze, models with \ce{CO} as the carbon carrier are required to explain the difference between the two \spitzer\ bands. The low inferred planet core mass is required to explain the large difference between the optical and NIR transit depths, as higher mass cores would result in reduced atmospheric scale heights.

    \item \textit{Spots + Photochemical Hazes: } A combined fit offers only marginal improvement from the spot-free haze fits. For tholin hazes, the improvement is not significant, and for soot hazes, the spot fraction required is inconsistent with the stellar spectrum and the fit to the transmission spectrum is still poor (high $\chi^2$) compared to the tholin-only fit. It is, however, encouraging that the resulting spot properties when using the tholin model are consistent with the two-temperature fit to the stellar spectrum. Adding spots does not change any of our conclusions with respect to the \ce{CO} vs \ce{CH4} models or low planet mass. 
\end{itemize}

The complication to our current analysis is that spot coverage fractions may change with time. The \ktwo\ and spectroscopy data were used to constrain the spots, but \ktwo\ data predate the other sets of data by $\sim$ 1.5 years. However, removing the \kepler\ data did not change the conclusions and the improvement to the fit was marginal. The IGRINS spectra were taken contemporaneously with the \mearth\ data and multiple spectra taken over several months (2016 Jan 25 to 2016 Mar 28) yield consistent results. There is also no evidence of extreme spot evolution in the long-term \mearth\ monitoring. Thus, these kinds of long-term effects are unlikely to change our findings, but may explain some of the tension in the tholin-haze fit.

A possibility we did not explore was a transiting ringed system \citep{Mamajek2012}. \cite{Ohno2022} highlight that even close-in planets can retain thick rings at young ($<100$\,Myr) ages, and rings would help explain the larger radii observed for young planets. Rings on K2-33\,b would need to be dominated by $\simeq$micron-sized dust and extend to at least twice the planetary radius to explain the deeper optical transits. Depending on the orientation and size distribution, rings would dominate over any signal from the planet's atmosphere \citep{Alam2022}, which would explain the lack of spectral features seen here. A companion paper, \citet{Ohno_inprep}, investigates this possibility further.

We conclude from our analysis that K2-33b likely hosts tholin-like hazes, with modest ($<10\%$) spot coverage for its host star. This scenario can explain all the data, including the light curve, stellar spectrum, and planetary transmission spectrum. More detailed information, including better constraints on the spot properties and haze production rate, will require more data. The current \hst\ data is particularly limiting, since it was only half a transit. A full transit of \textit{HST}/WFC3 combined with contemporaneous ground-based transit and spectroscopic monitoring would let us fully disentangle the effects of spot surface inhomogeneities from haze properties in the observed transmission spectrum. These observations will break this degeneracy because stellar spots generally increase the strength of the NIR H$_2$O feature, while hazes generally weaken it. Further, the former changes the observed out-of-transit spectrum while the latter does not, and each scenario will have a significantly different impact on the transit depths at bluer wavelengths depending on the spot and haze properties. Contemporaneous data would also resolve issues of spot properties changing with time and enable us to do a simultaneous fit of all constraints instead of the one-at-a-time fit we did here. 

Our haze results are in line with those of \citet{ohno2020}, which suggested that moderate haze production rates combined with high values of the eddy diffusion coefficient lead to super-Rayleigh optical slopes in transmission spectra. While both of these parameters are loosely constrained in exoplanet atmospheres, the high eddy diffusion coefficient may be reasonable given the young age and therefore high luminosity of K2-33b, and potentially tying into the significant upwelling of \ce{CO}-rich gas suggested by the \textit{Spitzer} data. Our results also fit those of \citet{Gao2020}, which predicted significant enlargement of young planet transit radii due to high altitude hazes entrained in atmospheric outflows, though here the effect seems to impact the optical wavelengths much more than the NIR.  Both the core mass and atmospheric mass fraction implied by our results are similar to those of typical sub-Neptunes \citep{Owen2017}, suggesting that K2-33b will eventually join that population after thermal evolution and contraction. 

The need for \ce{CO} to be the carbon carrier in our best-fit models (regardless of spot levels) is extremely compelling, but it relies entirely on the \spitzer\ bands. While the difference in depth holds even when we try different analysis methods, the significance of the difference changes depending on how we handle combining multiple transits (the difference is not significant with a single transit). 

Upward mixing of \ce{CO} is only one way in which it can supersede \ce{CH4} as the dominant carbon carrier in an atmosphere at the temperature of K2-33b. Alternatively, K2-33b's atmosphere can possess a low C/O and an enhanced metallicity, which would provide clues to its formation and early evolution \citep[e.g.,][]{Oberg2011, Madhusudhan2012}. Due to the large scale heights required to match the observed transmission spectrum, the metallicity cannot be too high (e.g. $>$100 $\times$ solar). Detecting a water and/or \ce{CO2} feature would help differentiate between these scenarios. We eagerly await confirmation of this detection with \textit{HST} and/or \textit{JWST}, which would enable a more precise measurement of the carbon-to-oxygen ratio.

\acknowledgements

We thank the anonymous referee for their careful reading and thoughtful comments on the manuscript.

The authors would like to thank Laura Kreidberg and Alex Teachey for helpful conversations regarding the \hst\ reduction. 

The authors also wish to acknowledge Wally, Penny, and Bandit for their input, sound counsel, and their constant quest to discover the unknown. 

P.C.T was supported by NSF Graduate Research Fellowship (DGE-1650116), the NC Space Grant Graduate Research Fellowship, the Zonta International Amelia Earhart Fellowship, and the Jack Kent Cooke Foundation Graduate Scholarship. This work was made possible by grants to A.W.M. from the {\it K2} Guest Observer Program (80NSSC19K0097) and  NASA's Astrophysics Data Analysis Program (80NSSC19K0583). M.J.F was supported by the NC Space Grant Graduate Research program and a grant from NASA's Exoplanet Research Program (80NSSC21K0393). 

This paper includes data collected by the K2 mission. Funding for the K2 mission is provided by the NASA Science Mission directorate.

This work is based on observations made with the \spitzer\ \textit{Space Telescope}, which is operated by the Jet Propulsion Laboratory, California Institute of Technology under a contract with NASA. Support for this work was provided by NASA through an award issued by JPL/Caltech. 

This research is based on observations made with the NASA/ESA Hubble Space Telescope obtained from the Space Telescope Science Institute, which is operated by the Association of Universities for Research in Astronomy, Inc., under NASA contract NAS 5–26555. These observations are associated with program ID: 14887 (PI: Benneke).

The MEarth Team gratefully acknowledges funding from the David and Lucile Packard Fellowship for Science and Engineering (awarded to D.C.). This material is based upon work supported by the National Science Foundation under grant AST-1616624, and work supported by the National Aeronautics and Space Administration under Grant No. 80NSSC18K0476 issued through the XRP Program.

\software{\texttt{LDTK} \citep{2015MNRAS.453.3821P}, \texttt{Iraclis} , \texttt{POET} + BLISS \citep{StevensonTransit2012}, \texttt{MISTTBORN} , \texttt{emcee}  \citep{Foreman-Mackey2013} , \texttt{corner.py} \citep{foreman2016corner}, \texttt{celerite} \citep{celerite}, matplotlib \citep{hunter2007matplotlib}, \texttt{batman} \citep{Kreidberg2015}, \texttt{Fleck} \citep{Morris2020}, \texttt{Astropy} \citep{astropy:2013, astropy:2018}, \texttt{numpy} \citep{harris2020array}}

\facilities{\spitzer\ (IRAC), \ktwo\ , \mearth\ , \hubble\ (WFC3)}

\clearpage
\bibliography{bib.bib,pg.bib}
\end{document}